\documentclass[10pt,journal,final,twocolumn]{IEEEtran}
\usepackage{amsfonts}
\usepackage{amsmath,amssymb}
\usepackage{graphicx}
\usepackage{color, soul}
\usepackage{algorithm,algorithmic}
\usepackage{bm}
\usepackage{booktabs}
\usepackage{flushend}
\usepackage{tikz}
\usepackage{cite}
\usepackage{url}
\usetikzlibrary{arrows}
\usepackage{subfigure}
\usepackage{indentfirst}
\usepackage[amsmath,thmmarks]{ntheorem}
\usepackage{theorem}
\usepackage{acronym}  
\usepackage{balance}

\newtheorem{lemma}{\bf Lemma}

\newtheorem{remark}{\bf Remark}

\theoremheaderfont{~~~\it}\theorembodyfont{\upshape}%
\theoremstyle{nonumberplain}
\theoremseparator{}
\theoremsymbol{\rule{1ex}{1ex}}

\acrodef{EAR}[EAR]{element activation ratio}
\acrodef{SNR}[SNR]{signal-to-noise ratio}
\acrodef{SINR}[SINR]{signal-to-interference-plus-noise ratio}
\acrodef{TC}[TC]{transmission coefficients}
\acrodef{US-RIS}[US-RIS]{user-specific RIS}
\acrodef{BSS-RIS}[BSS-RIS]{base-station-specific RIS}
\acrodef{DoF}[DoF]{degree of freedom}
\acrodef{FPGA}[FPGA]{field programmable gate array}
\acrodef{RF}[RF]{radio-frequency}
\acrodef{RIS}[RIS]{reconfigurable intelligent surfaces}
\acrodef{UE}[UE]{user equipment}
\acrodef{DL}[DL]{downlink}
\acrodef{TA}[TA]{transmit antenna}
\acrodef{RA}[RA]{receive antenna}
\acrodef{LoS}[LoS]{line-of-sight}
\acrodef{UL-TBF}[UL-TBF]{uplink transmit beamforming}
\acrodef{TPS}[TPS]{transmit phase shifter}
\acrodef{RC}[RC]{receiver combining}

\def \H {^H}
\def \opt {^{\text{opt}}}
\def \v {\bm v}
\def \w {\bm w}
\def \g {\bm g}
\def \f {\bm f}
\def \T {\bm \Theta}
\def \t {\bm \theta}
\def \x {\bm \xi}
\def \Pmax {P_{\text{max}}}
\def \ml {multi-layer }
\def \tb {transmit beamformer }
\def \sl {single-layer }
\newcommand{\RNum}[1]{\uppercase\expandafter{\romannumeral #1\relax}}
\ifCLASSINFOpdf
\else
\fi
\begin{document}
\title{Compact User-Specific Reconfigurable Intelligent Surfaces for Uplink Transmission}
\author{{Kunzan Liu, Zijian Zhang, Linglong Dai, and Lajos Hanzo
\vspace*{-1em}}
\thanks{Part of this work has been accepted by IEEE Global Communications Conference (IEEE GLOBECOM’21) \cite{liu2021user}.}
\thanks{This work was supported in part by the National Key Research and Development Program of China (Grant No. 2020YFB1807201), in part by the National Natural Science Foundation of China (Grant No. 62031019), and in part by the European Commission through the H2020-MSCA-ITN META WIRELESS Research Project (Grant No. 956256). L. Hanzo would like to acknowledge the financial support of the 
Engineering and Physical Sciences Research Council projects EP/P034284/1 
and EP/P003990/1 (COALESCE) as well as of the European Research 
Council's Advanced Fellow Grant QuantCom (Grant No. 789028). {\it (Corresponding author: Linglong Dai.)}}
\thanks{K. Liu, Z. Zhang, and L. Dai are with the Beijing National Research Center for Information Science and Technology (BNRist) as well as the Department of Electronic Engineering, Tsinghua University, Beijing 100084, China (e-mails: lkz18@mails.tsinghua.edu.cn, zhangzj20@mails.tsinghua.edu.cn, daill@tsinghua.edu.cn).}
\thanks{L. Hanzo is with the Department of Electronics and Computer Science,
University of Southampton, Southampton SO17 1BJ, U.K. (e-mail: lh@ecs.soton.ac.uk).}
}

\maketitle

\begin{abstract}
Large-scale antenna arrays employed by the base station (BS) constitute an essential next-generation communications technique.
However, due to the constraints of size, cost, and power consumption, it is usually
considered unrealistic to use a large-scale antenna array at the user side.
Inspired by the emerging technique of reconfigurable intelligent surfaces (RIS), we firstly propose the concept of user-specific RIS (US-RIS)
for facilitating the employment of a large-scale antenna array at the user side in a cost- and energy-efficient way.
In contrast to the existing employments of RIS, which belong to the family of
base-station-specific RISs (BSS-RISs), the US-RIS concept by definition facilitates the employment of RIS at the user side for the first time.
This is achieved by conceiving a \ml structure to realize a compact form-factor.
Furthermore, our theoretical results demonstrate that,
in contrast to the existing \sl structure, where only the phase of the signal reflected from RIS can be adjusted, the amplitude of the signal penetrating \ml US-RIS can also be partially controlled,
which brings about a new degree of freedom (DoF) for beamformer design that can be beneficially exploited for performance enhancement.
In addition, based on the proposed multi-layer US-RIS, we formulate the signal-to-noise ratio (SNR)
maximization problem of US-RIS-aided communications.
Due to the non-convexity of the problem introduced by this \ml structure,
we propose a \ml transmit beamformer design relying on an iterative algorithm
for finding the optimal solution
by alternately updating each variable.
Finally, our simulation results verify
the superiority of the proposed multi-layer US-RIS
as a compact realization of a large-scale antenna array at the user side for uplink transmission.
\end{abstract}



\section{Introduction}
\label{Introduction}
The internet of things (IoT) has attracted much research interest in the communication community. In next-generation networks, the number of access points may reach millions per square kilometer in tele-traffic hotspots \cite{giordani2020toward}. The interactions among smart devices, such as robots, intelligent reconfigurable furniture, and vehicles will require increasing wireless communications quality. In downlink transmissions, massive multiple-input multiple-output (MIMO) systems boost the capacity by employing a large-scale array having hundreds of antennas at the base station (BS) side \cite{rusek2012scaling}.
It has been verified that massive MIMO systems are capable of achieving orders of magnitude increase in spectral efficiency by simultaneously serving a massive number of users \cite{marzetta2010noncooperative}. 

However, in uplink transmissions, employing a large-scale array at the user side has been considered as unrealistic.
Although a large-scale beamformer generates high-gain beams, which increases the channel capacity and enhances the wireless coverage,
there is a fundamental dimensionality limit that prevents this idea.
Specifically, traditional antenna arrays require numerous \ac{RF} components, like \ac{RF} chains in the fully digital architecture \cite{el2014spatially} and phased arrays in the hybrid architecture \cite{letaief2019roadmap}. Hence this would lead to bulky circuits, high hardware cost, and excessive power consumption at the user side.

As a remedy, the emerging innovative technique of \ac{RIS} constituted by a large-scale array developed from meta-materials may be harnessed \cite{hu2018beyond,pan2021reconfigurable}. 
The appeal of RISs is that they can enhance the communications in a cost- and energy-efficient way, which provides us with a promising opportunity to construct large-scale arrays in a compact form at the user side.

\subsection{Prior contributions}
Again, a RIS is a large-scale array composed of a large number of passive low-cost elements, which can reflect or transmit the incident electromagnetic waves in the desired directions by appropriately adjusting their phase shifts \cite{basar2019wireless,liu2021star}.
In contrast to the traditional energy-hungry \ac{RF} components (e.g. 250\,mW per RF chain) \cite{amadori2015low},
the \ac{RIS} elements are passive, and yet they can engender the required phase shifts in a relatively energy-efficient way \cite{liaskos2018using,bai2020latency,hu2021robust,fang2020energy,yu2021irs}.
Since the benefits of RISs have already been verified in practical wireless communications prototypes \cite{dai2020reconfigurable,tang2019wireless},
\acp{RIS} have been regarded as a promising technique
for future 6G wireless communications \cite{liang2019large}.



The applications of RISs can be generally divided into two broad categories.
The first category employs relay-like RISs between the BS and the user for enhancing the transmission links.
For example, RISs can provide additional propagation paths and thus can be utilized for improving the transmission reliability by overcoming blockages \cite{basar2019wireless}.
Another scenario is considered where multiple BSs and multiple RISs are employed for simultaneously serving multiple users. As a benefit, compared to traditional networks operating without RISs, an improved network capacity and energy efficiency can be achieved \cite{Zhang'20,liu2021energy,huang2019reconfigurable}.
In \cite{wu2019intelligent}, RISs are used for reducing the transmit power of the BS, which is derived by jointly optimizing the phase shifts of RIS elements as well as the parameters of both the BSs and users.
As for the placement of the relay-like RIS, You \textit{et al.} \cite{you2020how} has investigated the RIS deployment strategies for approaching optimal performance\footnote{Note that, although \cite{you2020how} investigated the practicability of employing relay-like RISs close to the user, which seems similar to the proposed concept of user-specific RIS (US-RIS) in Subsection~\ref{Concept}, it can still be categorized as a base-station-specific RIS (BSS-RIS). The differences between these two contributions are explicitly detailed in Subsection~\ref{Architecture}, Table~\RNum{1}.}.

The second category employs RISs near the BS for analog beamforming
at a low cost and frugal power consumption.
The authors of \cite{lu2020reconfigurable} propose a RIS-based transmit precoding architecture for maximizing the sum-rate,
where the phased arrays are replaced by the RIS at the BS side.
As a further development, the authors of \cite{yang2020beamforming} utilize a RIS at the BS side for beamforming relying on the optimal phase shifts.
In these contributions, RISs serve as part of 
a new BS type that transmits/receives signals at a reduced cost and power consumption \cite{lu2020reconfigurable,yang2020beamforming}.

In a nutshell, both the above two categories utilize RISs for improving the communication performance of the cellular network, hence we refer to them as \acp{BSS-RIS}.
By contrast, RISs have not been employed at the user side in the open literature.

\subsection{Our contributions}
To circumvent the dimensionality limit of employing a large-scale array
at the user side, we propose the concept of \acp{US-RIS}, where the RIS is employed at the user side for enhancing the user's capability in uplink transmissions\footnote{Simulation codes are provided to reproduce the results in this paper: \url{http://oa.ee.tsinghua.edu.cn/dailinglong/publications/publications.html}.}.
Specifically, the contributions of this paper can be summarized as follows.
\begin{itemize}
\item
We propose the concept of \ac{US-RIS}
to break the dimensionality limit and to facilitate the employment of a large-scale array at the user side.
In contrast to the existing \acp{BSS-RIS}, which are utilized for improving the communication performance of the cellular network, this is the first use case of RIS dedicated to the user side.
\item
We propose and analyze a \ml structure to realize \acp{US-RIS} by considering the space-limited characteristics of users.
Explicitly, in contrast to existing RISs, where only the phase of the signal can be adjusted, the amplitude of the signal can also be partially controlled.
This provides us with an additional \ac{DoF} for RIS beamforming design.
\item
We then formulate the associated \ac{SNR} maximization problem of
US-RIS-aided communications.
Due to the non-convexity introduced by the \ml RIS structure proposed,
we develop a \ml transmit beamformer design
that can control both the phase and the amplitude of the transmitted signal.
Specifically, the proposed \ml transmit beamformer design relies on an
iterative optimization algorithm for finding the optimal solution to the problem formulated.
Finally, our simulation results verify the benefits of the proposed \ml US-RIS as a realization of a large-scale array at the user side for uplink transmission.
\end{itemize}

\subsection{Organization and notation}

\textit{Organization:}
The rest of the paper is organized as follows.
In Section~\ref{System Model},
we propose the basic concept of US-RIS.
Then, we conceive a novel architecture relying on this multi-layer US-RIS.
In Section~\ref{Performance Analysis},
we theoretically analyze the performance of the proposed multi-layer structure.
In Section~\ref{Precoding-Design},
we consider a US-RIS-aided communications scheme
and formulate the corresponding SNR maximization problem.
The \ml transmit beamformer design is proposed as an iterative algorithm for solving the problem formulated, where each variable is updated in an alternating fashion.
In Section~\ref{Simulation Results},
simulation results are provided
for quantifying the performance of US-RIS-aided communications,
demonstrating the practical realization of a large-scale array at the user side for uplink transmission.
Finally, in Section~\ref{Conclusions}, we provide our conclusions followed by promising future research ideas.

\textit{Notation:} $\mathbb C, \mathbb R,$ and $\mathbb R_+$ denote the set of
complex, real, and positive real numbers, respectively;
$[L]$ represents the set of integers $\{1,2,\cdots,L\}$;
$\bm A^{-1}, \bm A^*,\bm A^T,$ and $\bm A^H$ denote the inverse, conjugate, transpose, and conjugate transpose of matrix $\bm A$, respectively; 
$\Vert\bm x\Vert_{2}$ is the $\ell_{2}$-norm of vector $\bm x$; 
$\left\langle\bm x\right\rangle$ is the normalized vector of $\bm x$, i.e., $\left\langle\bm x\right\rangle=\bm x/\Vert\bm x\Vert_{2}$;
$\text{arg}(\bm x)$ and $\text{exp}(\bm x)$ denote the phase angle and exponential representation of each element of the complex vector $\bm x$, respectively;
$\vert x\vert$ denotes the amplitude
of a complex scalar $x$; 
$\text{diag}(\cdot )$ is the diagonal operation;
$\mathcal{CN}\left(\bm \mu, \bm \Sigma \right)$ represents the complex multivariate Gaussian distribution with the mean $\bm \mu$ and the variance $\bm \Sigma$;
$\bm 0_{L}$ denotes the $L\times 1$ zero vector;
$\bm I_{L}$ denotes the $L\times L$ identity matrix;
$\bm \psi_{\text{max}}(\bm A)$ is the eigenvector of matrix
$\bm A$ corresponding to its largest eigenvalue.

\section{System Model}
\label{System Model}
To realize a large-scale array
at the user side at a low cost and low power consumption,
in this section, we propose the concept of US-RISs. 
Specifically, the concept is introduced in Subsection~\ref{Concept}.
Then, we propose a novel architecture relying on a multi-layer US-RIS and compare it to existing BSS-RISs in Subsection~\ref{Architecture}.
Finally, in Subsection~\ref{US-RIS-aided communications},
we develop the system model of US-RIS-aided
communications.

\subsection{Concept of US-RIS}
\label{Concept}
As the terminology suggests, US-RIS is a hardware technique conceived for constructing a large-scale array at the user side.

\begin{figure*}[!t]
\centering
	\includegraphics[trim = 0 190 0 190, clip, width=.95\textwidth]{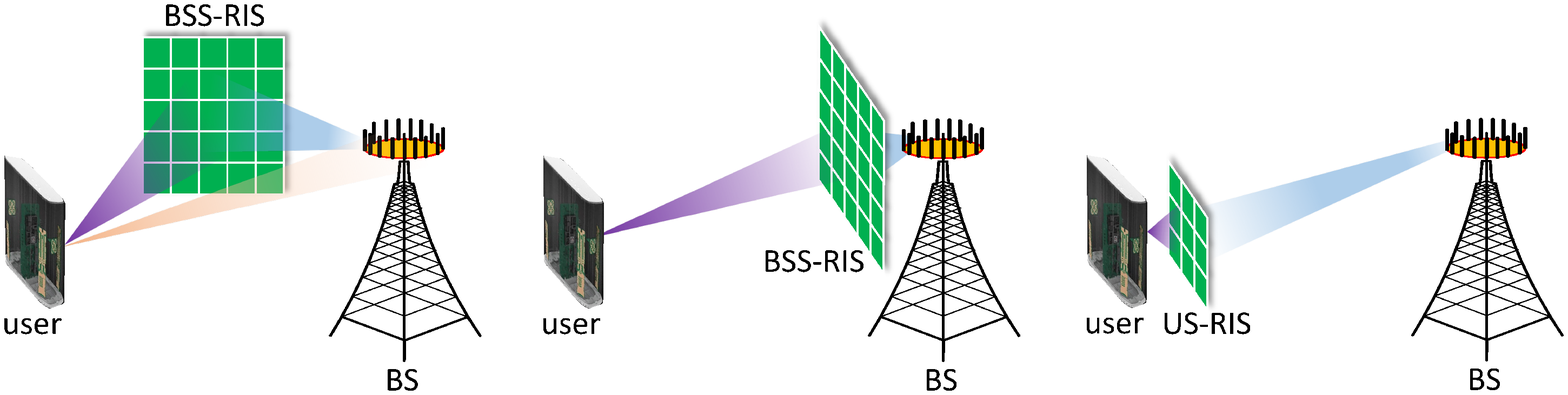}
	\centerline{(a)~~~~~~~~~~~~~~~~~~~~~~~~~~~~~~~~~~~~~~~~~~~~~~
	                  (b)~~~~~~~~~~~~~~~~~~~~~~~~~~~~~~~~~~~~~~~~~~~~~(c)}
	\caption{Different communication systems utilizing RIS.
		(a) BSS-RIS-aided communications, where a relay-like RIS is employed between the BS and the user for enhancing the multipath diversity by mitigating blockages.
		(b) BSS-RIS-aided communications, where the RIS is employed at the BS side.
		(c) US-RIS-aided communications, where the RIS is employed at the user side.}
	\label{system}
\end{figure*}

Again, traditional applications of RISs in wireless communications are limited to employing RISs between the BS and the user or alternatively, at the BS side, as shown in Fig. \ref{system} (a) and (b), respectively. In Fig. \ref{system} (a), a relay-like RIS is used to ``reconfigure'' the wireless channels for enhancing the multipath diversity by mitigating blockages. By contrast, in Fig. \ref{system} (b), a RIS is employed at the BS side for beamforming. 
Again, both of these two applications aim for improving the performance of the cellular network, which are termed as \acp{BSS-RIS}.

By contrast, here we conceive the US-RIS concept shown in Fig. \ref{system} (c). 
In contrast to the traditional \acp{BSS-RIS}, US-RISs are employed close to the user as a realization of a large-scale array at the user side.
Given their cost- and energy-efficient characteristics,
US-RISs are eminently suitable for compact user-side employment, which dispels the myth that a user can hardly harness a large-scale array, considering the inherent cost- and power-constraints.

\subsection{Architecture}
\label{Architecture}

\begin{figure}[!t]
\begin{minipage}{0.38\linewidth}
  \centerline{\includegraphics[trim = 0 0 0 0, clip, width=.95\linewidth]{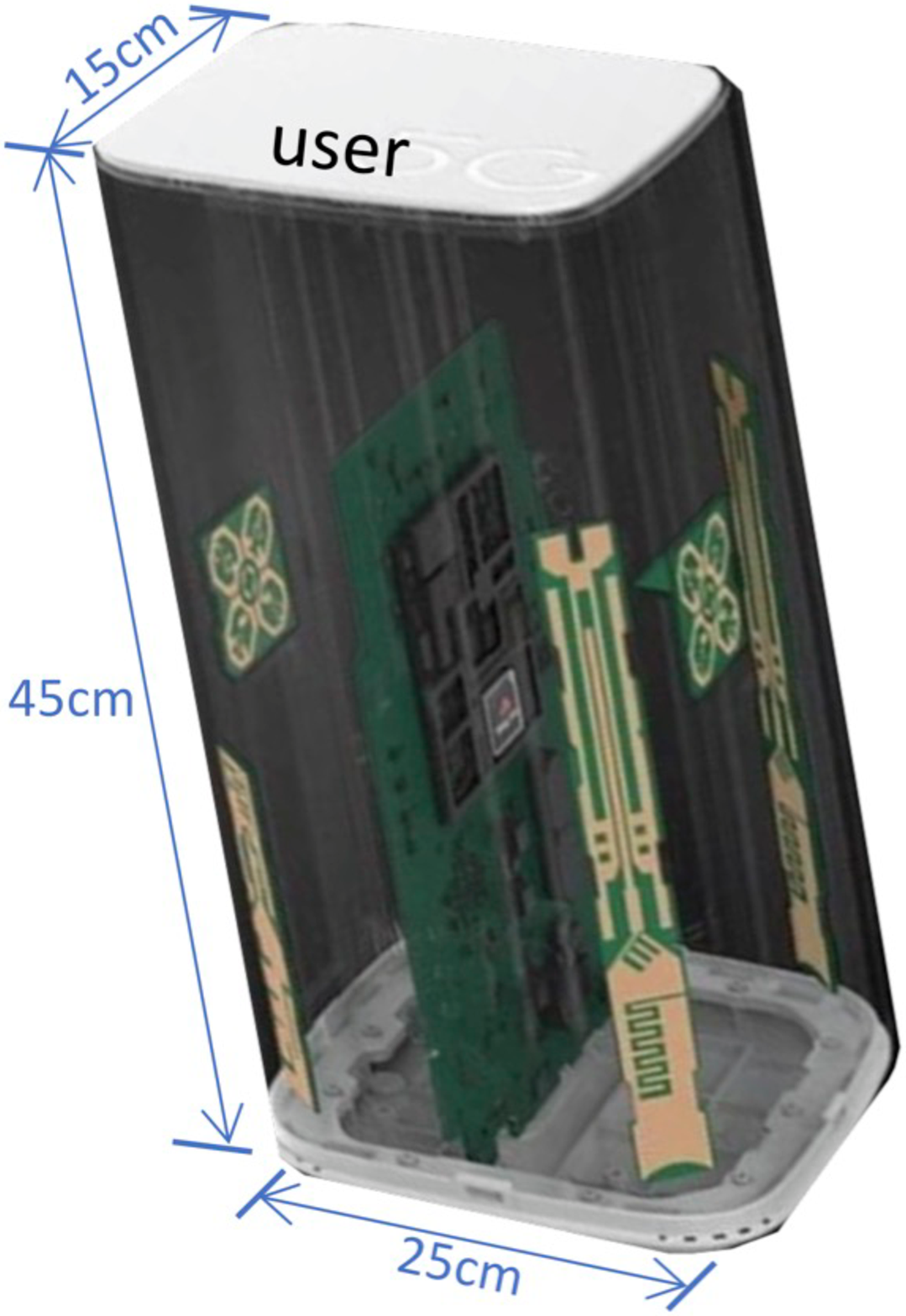}}
  \vspace{1.2em}
  \centerline{(a)}
\end{minipage}
\hfill
\begin{minipage}{0.62\linewidth}
  \centerline{\includegraphics[trim = 0 120 0 105 , clip, width=.882\linewidth]{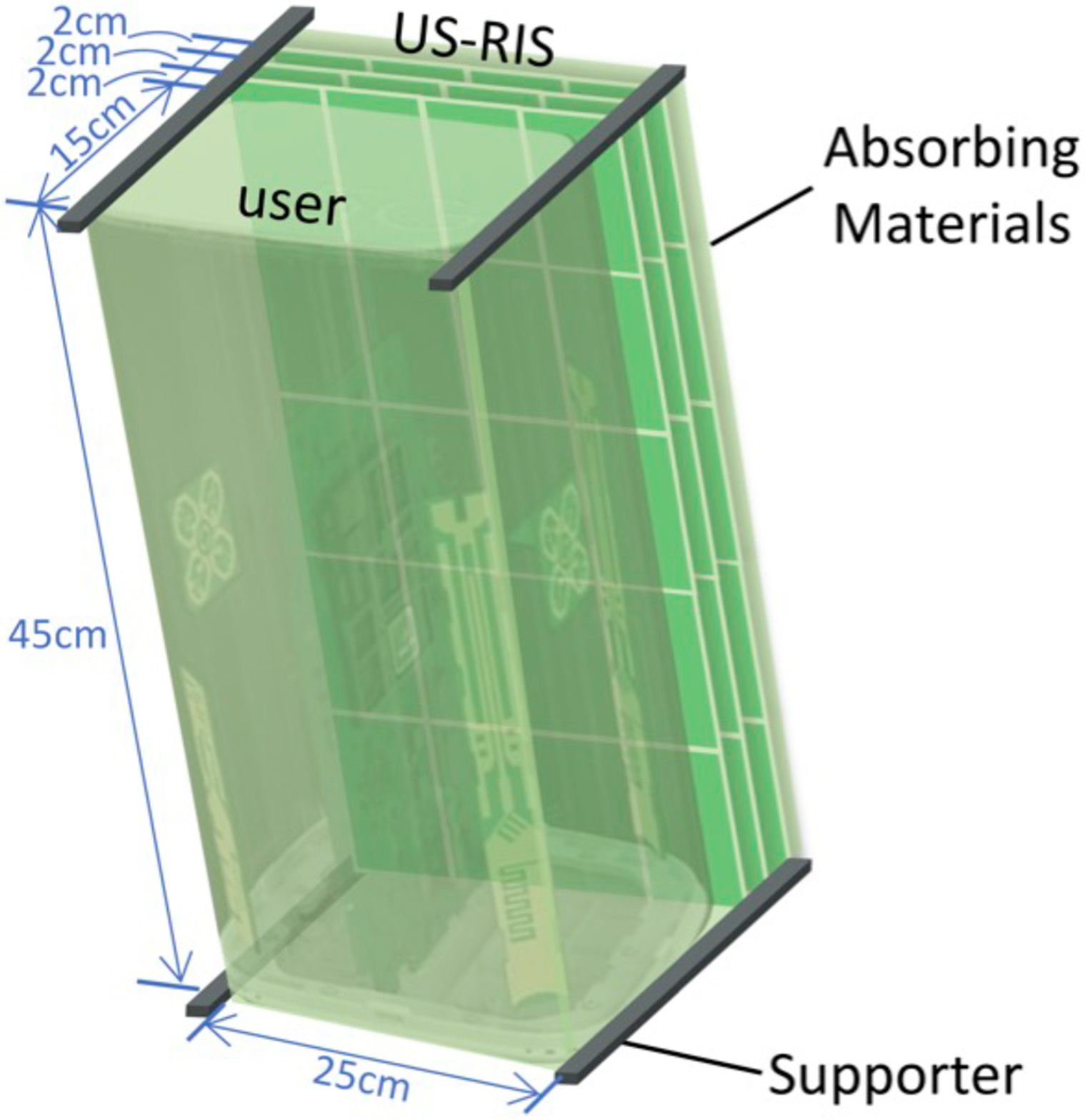}}
  \centerline{(b)}
  \end{minipage}
\caption{US-RIS construction. 
(a) A typical user which is a real-world model of customer-premises equipment (CPE).
(b) The proposed architecture with multi-layer US-RIS.}
\label{architecture}
\end{figure}

To illustrate a potential US-RIS construction,
Fig. \ref{architecture} gives an example based on a customer-premises equipment (CPE).
Specifically, this CPE facilitates access to communication routers, network switches, and networking adapters.
Fig. \ref{architecture} (a) portrays a real-world CPE model of size
$25\,\text{cm}\times15\,\text{cm}\times45\,\text{cm}$.
The physical size of this transmissive US-RIS is beneficially reduced by constructing a multi-layer structure.
Fig. \ref{architecture} (b) shows the axonometric construction of the US-RIS, where three layers of RISs are vertically stacked in front of the CPE in order to form a multi-layer structure, with a $2\,\text{cm}$ gap between adjacent vertical layers. 
In this way, the overall size of the resultant US-RIS-aided CPE becomes
$25\,\text{cm}\times21\,\text{cm}\times45\,\text{cm}$,
which may be deemed acceptable for a domestic installation.
Note that the number of layers and the inter-layer gaps are flexible in practical applications, depending on the dimensions of the user equipment, the trade-off between cost and performance, etc.
Naturally, the entire structure should be hosted by an enclosure, surrounded by absorbing materials to reduce the energy loss and to protect the internal channel from external interference. 
As usual, the controller is informed by the channel estimator on how to configure the phase shifts of US-RISs.
For practical hardware implementation, \cite{cui2020programmable} proposes a similar construction for implementing a deep neural network (DNN), which may be adopted for realizing this architecture.

\begin{table}[t] \vspace{0.6em}
\caption{Comparison between BSS-RIS and US-RIS}
\label{CharacteristicTable}
\centering
\begin{tabular}{|l|r|r|}
    \hline
    \textbf{Aspect} & \textbf{BSS-RIS} & \textbf{US-RIS} \\\hline
		Location& BS& user\\
Controller& one/multiple BS(s)& one user\\
Beneficiary& ~~one/multiple user(s)& one user\\
Operating mode~~& mainly reflective& transmissive\\
Structure&single-layer&~~~~~single/multi-layer\\
Size& large& compact\\
EAR& small& big\\\hline
    \end{tabular}
\end{table}

To clarify the characteristics of this novel user-side architecture and the differences between BSS-RIS and US-RIS, we summarize their key aspects in Tab. \ref{CharacteristicTable} and compare them as follows.

\begin{itemize}
\item
\emph{Location, controller, and beneficiary:}
According to their terminologies,
the most essential difference between
BSS-RIS and US-RIS is their location.
The BSS-RIS is primarily proposed for improving the overall channel capacity of the cellular network, while US-RIS at the user side is transmissive and may also be regarded as an integral component of the user.
Moreover, it has been theoretically shown that due to the intrinsic “multiplicative fading” effect of passive RIS \cite{najafi2020physics,zhang2021active}, the RIS is preferred to be employed near the BS or the user to minimize the pathloss \cite{wu2021intelligent,you2021wireless}.
Thus, we place the US-RIS near the user is in line with the results of RIS placement optimization.
Also, both the controller and the beneficiary of the BSS-RIS and US-RIS are different.
The BSS-RIS is controlled by one/multiple BS(s) in the system
and it cooperatively serves one/multiple user(s) at the same time.
By contrast, for the US-RIS, both the controller and the beneficiary is the same user. 
Both the beamforming design and phase-shift control are carried out at the US-RIS.
\item
\emph{Operating mode and structure:}
Most BSS-RISs tend to be reflective arrays \cite{Zhang'20,liu2021energy,wu2019intelligent,lu2020reconfigurable}, while our US-RIS is a transmissive array operating under a tight space constraint at the user side \cite{zeng2021reconfigurable}.
For US-RIS, the multi-layer structure is conceived for further improving the array gains attained in a limited space. 
To elaborate, in contrast to the existing single-layer structure, which can only adjust the phase of the incident signal, our multi-layer structure has the advantage that it facilitates partial amplitude control of the radiated signals.
This feature provides a new \ac{DoF} for beamforming design as a structural benefit, which will be detailed in Section~\ref{Performance Analysis}.
\item
\emph{Size and \ac{EAR}:}
In BSS-RIS-aided communications,
the size of the BSS-RIS is expected to be very large, which is not practical for the user.
To limit the user equipment's dimension and power consumption, each layer of the US-RIS should be much smaller than a BSS-RIS. 
However, the proposed multi-layer structure has a compact size. 
Concretely, the \ml RIS simultaneously controls the phase shifts of the different RIS layers.
We now proceed by introducing the new metric of element activation ratio (EAR), which explicitly quantifies the particular fraction of activated elements of a transmissive RIS.
Specifically, in this ratio, we count an element as being “activated”, if its power is higher than a threshold percentage $\varepsilon$ of the average power\footnote{Under this definition, a RIS will have an EAR 100\% with threshold percentage $\varepsilon$ when each element has a radiated power higher than $\varepsilon$ of the average power, while a RIS will have the minimum EAR, when a single element radiates all the power.}.
In Subsection~\ref{Analysis on RIS's element activation ratio},
we will provide experimental EAR results.
Apart from the compact structure of our US-RIS, a further particular benefit of the proposed multi-layer architecture is that it is capable of reducing the pilot overhead required for channel estimation, since the channels experienced by the parallel US-RIS layers are similar and can be estimated in advance using the methods of \cite{you2020wireless} for example.
\end{itemize}

\subsection{System model of US-RIS-aided communications}
\label{US-RIS-aided communications}

\begin{figure}[t] 
   \centering
   \includegraphics[width=.95\linewidth]{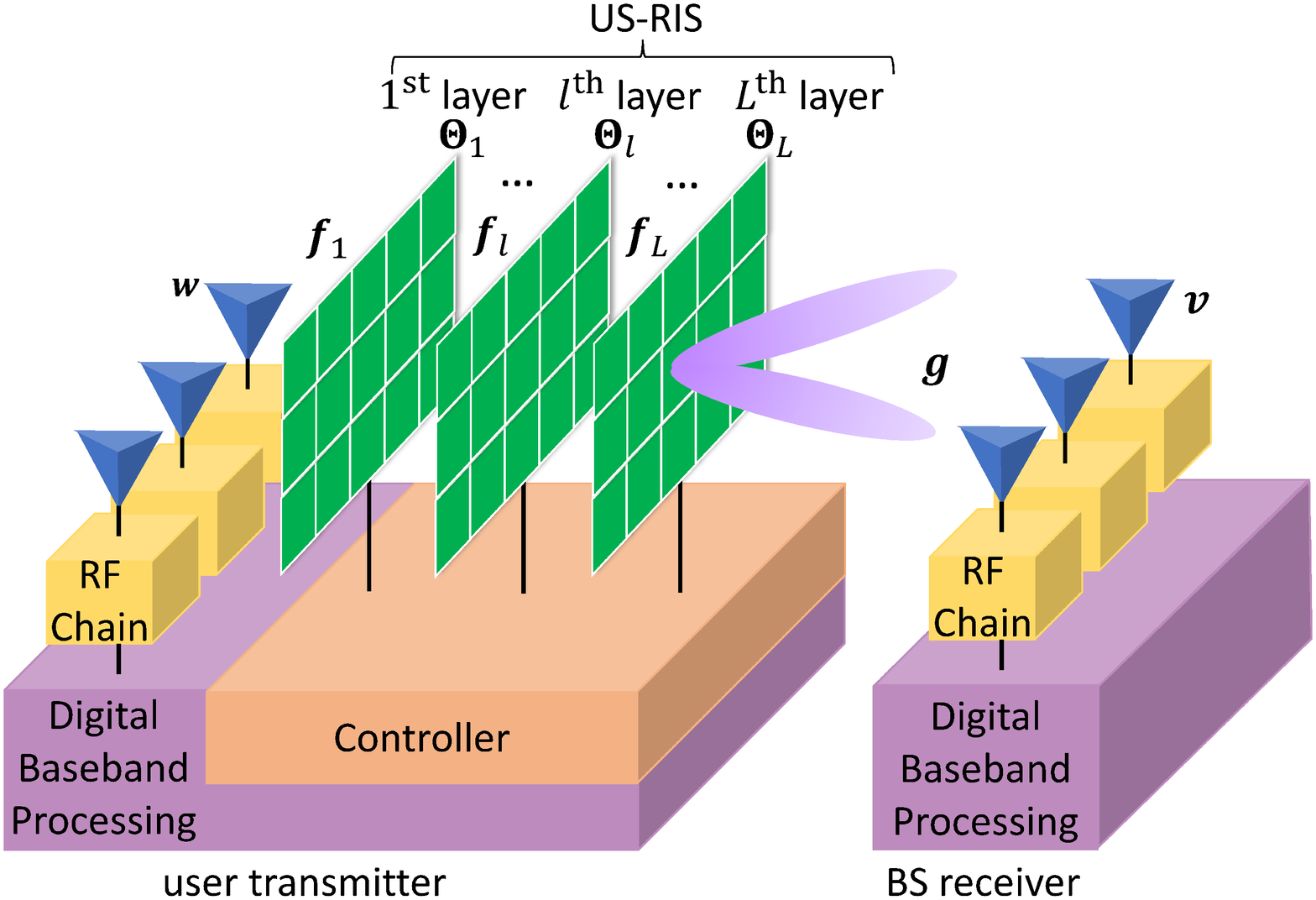} 
   \caption{System model of US-RIS-aided uplink transmission
   from a user to the BS, relying on the uplink transmit beamforming (UL-TBF) vector $\w$, transmit phase shifter (TPS) matrices $\T_{1},\cdots,\T_{L}$, and on the receiver combining (RC) vector $\bm v$ communicating over the wireless channels.}
   \label{MIMO}
\end{figure}

Let us consider the US-RIS-aided communications scenario of Fig. \ref{MIMO}, where a multi-layer US-RIS is used for enhancing the uplink transmission from the user to the BS.
Assume furthermore that the user and the BS are equipped with
$K$ \acp{TA} and $M$ \acp{RA}, respectively. 
The US-RIS is composed of $L$ layers having $N_l$ elements on the $l$-th layer. 
Without loss of generality, we assume\footnote{The \ml precoding design of US-RIS proposed in Section
	\ref{Precoding-Design} is unchanged even if the number of
	elements on each layer is different.
	However, it will affect the performance of communications.
	The discussion about the performance against the number of
	elements on each layer is left for future work.} that $N_l$ is equal to $N$ for all $l\in [L]$. 
Finally, we denote the $n$-th \ac{TA} element on the US-RIS's $l$-th layer as the $(l,n)$-th \ac{TA} element.

We denote the transmitted uplink symbol by $s\sim\mathcal{CN}\left(0,1\right)$.
The symbol $s$ is firstly processed by the \ac{UL-TBF} vector
$\w\in\mathbb C^{K\times 1}$ of Fig. \ref{MIMO}, subject to the power constraint
$\left\Vert\w\right\Vert_{2}^{2}\leq\Pmax$ before transmission.
Only the user-RIS-BS \ac{LoS} link is considered, while all other possible links are neglected.
To elaborate briefly on this assumption, since the user and the US-RIS are wrapped in a protecting enclosure, the transmitted signal must penetrate the multiple transmissive layers to be received at the BS\footnote{
Nonetheless, the user-BS link can also be assumed to exist, depending on the practical architecture. The proposed transmit precoder design can be readily extended to the scenario, where this user-BS link exists.
}.
Let us denote the \ac{TPS} matrix of US-RIS's $l$-th layer by
\begin{equation}
\label{PhaseShiftMatrix}
\T_{l}=\text{diag}\left(\t_{l}\right)
=\text{diag}\left(\left[\theta_{l,1},\cdots,\theta_{l,N}\right]^T\right),
\end{equation}
where $\theta_{l,n}\in\mathcal F$ of Fig. \ref{MIMO} is the phase shift
of the $(l,n)$-th element and $\mathcal F$ is the 
feasible set of the \ac{TC}. In this paper, we consider the widely used \ac{TC} set of
\begin{equation}
\label{FeasibleSet}
{{\mathcal F}} = \left\{ {\theta}{\Big |}\theta=e^{j\varphi},\varphi\in
\left[-\pi,\pi\right]\right\} ,
\end{equation}
which indicates that only the phase ${\theta _{l,n}}$ 
can be controlled independently and continuously \cite{wu2019intelligent}.
Therefore, the signal received by the BS can be modeled as
\begin{equation}
\label{ReceiveSignal}
\bm y=\g \H \left( \prod_{l=L}^{1}\kappa\T_{l}\f_{l}\right)\w s+\bm n,
\end{equation}
where $\kappa$ denotes the loss factor when the electromagnetic wave penetrates each layer.
Furthermore, $\g\in\mathbb C^{N\times M}, \f_{1}\in\mathbb C^{N\times K},
$ and $\f_{l}\in\mathbb C^{N\times N}$ of Fig. \ref{MIMO} denote the frequency-domain channels spanning from the US-RIS’s $L$-th layer to the BS, from the user to the US-RIS’s first layer, and from the US-RIS’s $(l-1)$-st layer to the $l$-th layer, for all $l\in\{2,3,\cdots,L\}$, respectively.
Still referring to \eqref{ReceiveSignal}, $\bm n\sim\mathcal{CN}\left(\bm 0_{M},\sigma^{2}\bm I_{M}\right)$ denotes the additive white Gaussian noise (AWGN) introduced at the BS receiver, where $\sigma^{2}$ is the noise power.
Finally, the BS receiver combines the
signal gleaned from $M$ antennas using a \ac{RC} vector $\v\in\mathbb C^{M\times 1}$ of Fig.~\ref{MIMO}.
Thus, the uplink signal combined at the BS receiver can
be represented as
\begin{equation}
\label{CombinedSignal}
z=\v\H\bm y
=\v\H\g \H \left( \prod_{l=L}^{1}\kappa\T_{l}\f_{l}\right)\w s+\v\H\bm n.
\end{equation}

To illustrate the benefits of the US-RIS based on this system model\footnote{The multi-layer structure can be applied at the user side or the BS side following a similar system model, and it can bring about the above-mentioned structural benefits at both sides.}, the theoretical analysis including the achievable SNR of our US-RIS-aided system will be provided in Section~\ref{Performance Analysis} and Section~\ref{Precoding-Design}, respectively.

\section{Performance Analysis on Multi-Layer Structure}
\label{Performance Analysis}

In this section, we provide further analysis and discussions on the firstly proposed multi-layer structure of RIS.
In Subsection~\ref{Theoretical analysis}, we theoretically analyze the performance of the proposed \ml US-RIS by considering a simplified system.
Then, in Subsection~\ref{Discussion on multi-RIS system}, we summarize and compare the existing contributions about the system with multiple cooperative RISs.

\subsection{Theoretical analysis}
\label{Theoretical analysis}

\begin{figure}[!t] 
   \centering
   \includegraphics[trim = 0 240 0 240, clip, width=.95\linewidth]{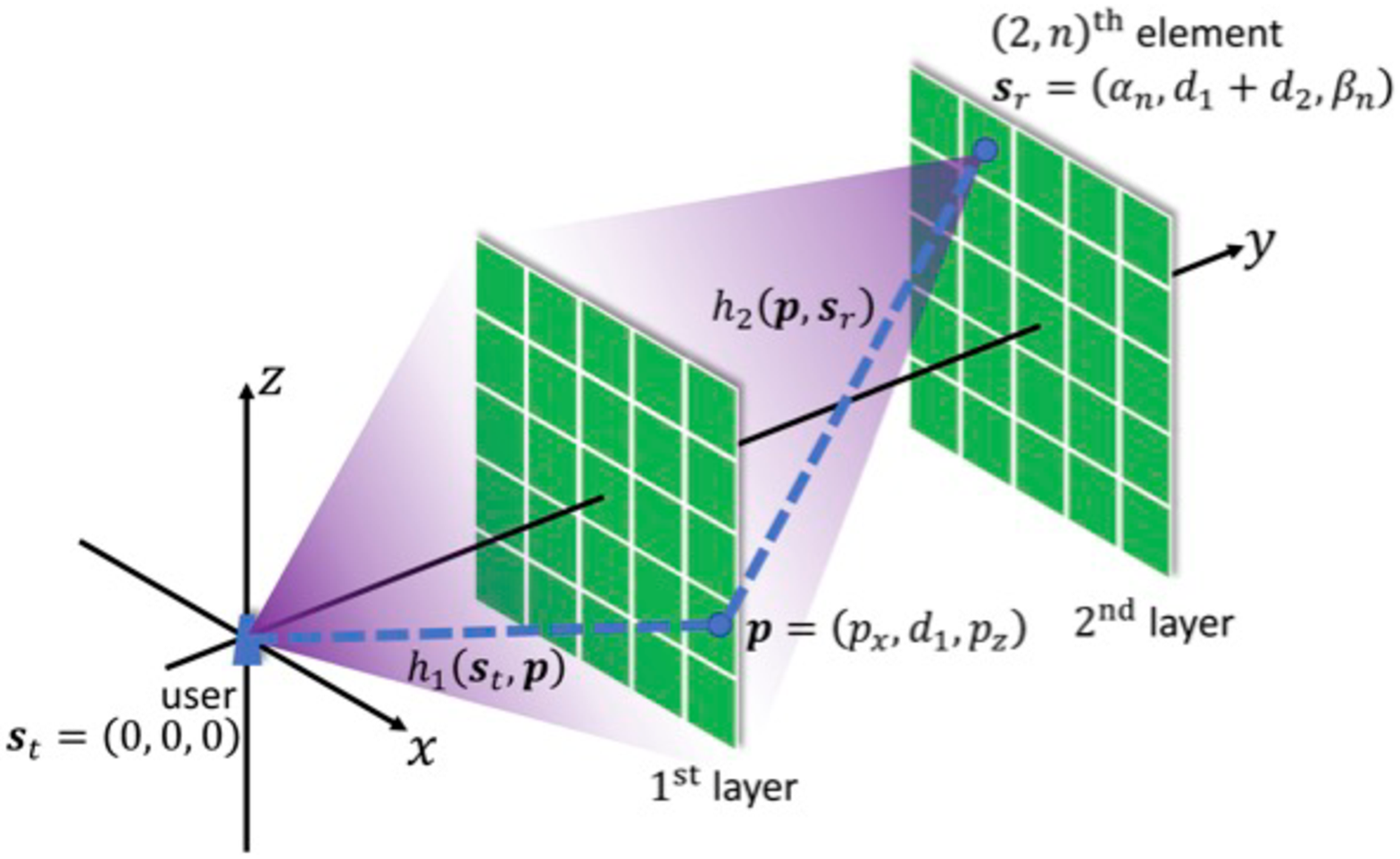} 
   \caption{A simplified system having $L=2$ layer is employed to assist
the uplink transmission from a single-antenna user.
The scale is exaggerated compared to a practical system
for clarity.}
   \label{theoretical}
\end{figure}

For the proposed multi-layer US-RIS, let us consider a simplified system where a multi-layer US-RIS having $L=2$ layers is employed to assist the uplink transmission from a single-antenna user, as shown in Fig. \ref{theoretical}.
Let us denote the number of elements in each layer by $N=b^{2}$, where each element is of size $a\times a$.
The position of the user is $\bm s_{t}=(0,0,0)$, and the US-RIS's two layers are positioned at the plane $y=d_{1}$ and $y=d_{1}+d_{2}$ respectively,
with their boundaries parallel to the coordinates and
their geometric centers on the $y$-axis.
Then, the position of each element is fixed.
Specifically, for the $(l,n)$-th element of US-RIS,
the position of the center can be written as
$\bm s_{r}=\left(\alpha_{n},\sum_{i=1}^{l}d_{i},\beta_{n}\right)$,
where $\alpha_{n}$ and $\beta_{n}$ are explicit functions with respect to $n$.
Thus, the region enclosed by the $(l,n)$-th element can be represented as
\begin{equation}
\label{region}
\Omega_{l,n}=\left\{
-\frac{a}{2}\leq x-\alpha_{n}\leq\frac{a}{2},
y=\sum_{i=1}^{l}d_{i},
-\frac{a}{2}\leq z-\beta_{n}\leq\frac{a}{2}
\right\}.
\end{equation}

\begin{figure}[!t] 
   \centering
   \includegraphics[trim = 230 30 220 35, clip, width=\linewidth]{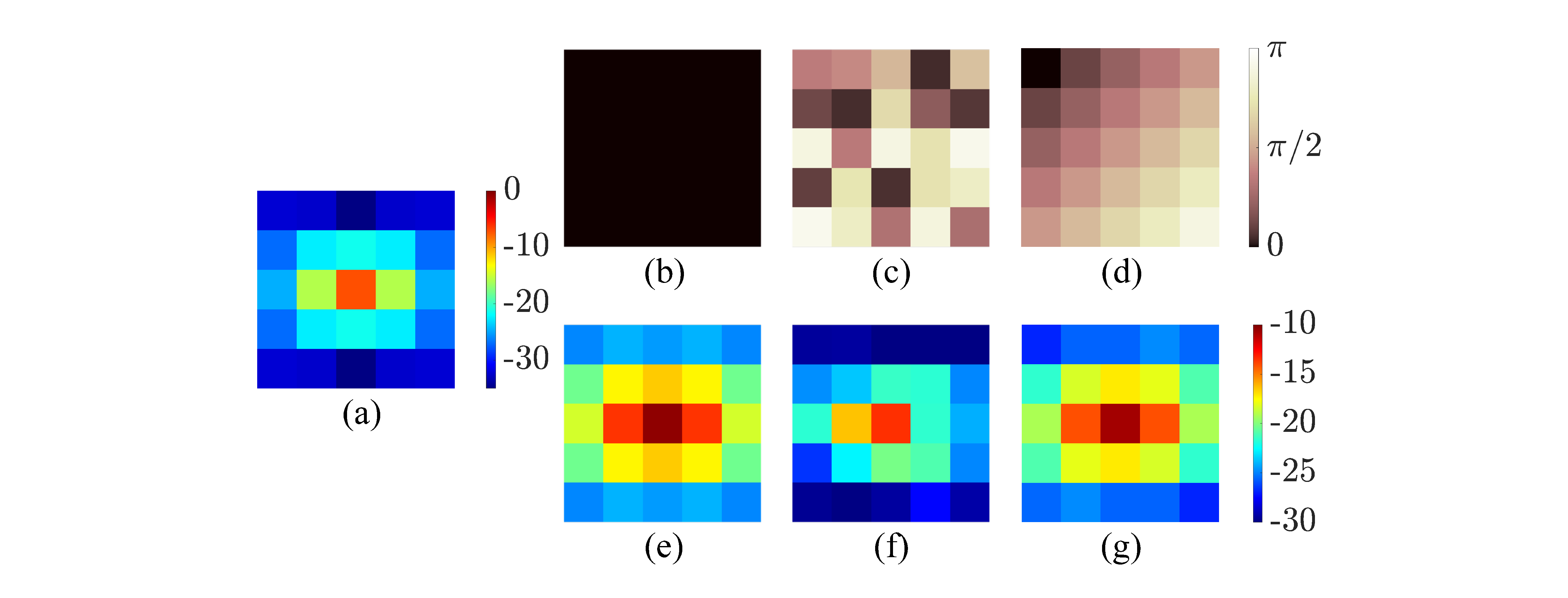} 
   \caption{An example of power distributions of the signal received on different layers for different phase-shift configurations. (a) Power distribution of the first layer (unit: dBW). (b)-(d) Three different phase shifts of the first layer of US-RIS (colorbar on the right is unified for 3 subfigures, unit: rad). (e)-(g) Power distributions of the second layer corresponding to the phase shifts displayed in (b)-(d) (colorbar on the right is unified for 3 subfigures, unit: dBW).}
   \label{DoFExample}
\end{figure}

We assume that, for single-layer RISs, only the phase shift of each element can be controlled continuously.
However, for the multi-layer structure, the power distribution of the second layer can be adjusted by controlling the phase shifts on the first layer, so that the radiated power can be focused on specific elements of the second layer.
This allows the amplitude of the signals penetrating \ml US-RISs to be controlled to some extent.

In order to understand this intuitively, a numerical example is given Fig. \ref{DoFExample}. 
Specifically, the power distribution of the signal received on the first layer is shown in Fig. \ref{DoFExample} (a). 
Note that, in contrast to the far-field case where the power is uniformly distributed, the power distribution shown in Fig. \ref{DoFExample} (a) is different, since the first layer lies close to the user, and thus the pathloss between the user and RIS elements at different positions is not the same.
Given the different phase-shift configurations on the first layer displayed in Fig. \ref{DoFExample} (b)-(d) (no phase-shift, random phase-shift, and gradual phase-shift, respectively), the patterns of power distributions on the second layer become different as shown in Fig. \ref{DoFExample} (e)-(g), indicating that the power distribution can be controlled by \ml precoding. 
Equivalently, it can be readily seen that the amplitude term of the signals penetrating the \ml US-RIS can be controlled to some extent, indicating that the a new DoF is granted for our US-RIS-aided transmit beamformer design.
The result is different from the far-field case, since when the power is uniformly distributed, pure phase-shift control of the RIS is sufficient for the beamformer.
Next, we formally state our theoretical result in \textbf{Lemma \ref{lem}}, and quantitative analysis will be provided in Subsection~\ref{Analysis on RIS's element activation ratio}.

\begin{lemma}
\label{lem}
Upon denoting the signal radiated from the $(2,n)$-th element of the US-RIS by $y_{n}$, the phase of $y_{n}$ can be adjusted to any desired angle
in $[-\pi,\pi]$, while the amplitude of $y_{n}$ can be adjusted in the range $\left[0,\zeta_{n}\right]$,
where we have:
\begin{equation}
\label{zeta_n}
\begin{aligned}
\zeta_{n}^{2}&=\frac{1}{16\pi^{2}}\int_{-\frac{ab}{2}}^{\frac{ab}{2}}
\emph{d}p_{x}
\int_{-\frac{ab}{2}}^{\frac{ab}{2}}
\emph{d}p_{z}\\
&\frac{d_{1}d_{2}\left(p_{x}^{2}+d_{1}^{2}\right)
\left[\left(p_{x}-\alpha_{n}\right)^{2}+d_{2}^{2}\right]}{
\left(p_{x}^{2}+p_{z}^{2}+d_{1}^{2}\right)^{5/2}
\left[\left(p_{x}-\alpha_{n}\right)^{2}+\left(p_{z}-\beta_{n}\right)^{2}+d_{2}^{2}\right]^{5/2}
}.
\end{aligned}
\end{equation}
\end{lemma}
\begin{IEEEproof}
\label{proof}
The proof is given in Appendix \ref{Method} in two steps. 
In particular, the closed-form expression of
$y_{n}$ is given in \eqref{ClosedForm},
and the legitimate ranges of the phase and amplitude are
then discussed.
\end{IEEEproof}

\begin{remark}
\label{remark}
In \textbf{Lemma \ref{lem}}, we have stated that, for a specific $n$, the phase and amplitude of $y_{n}$ can be controlled in a range.
However, due to the coupling of $\theta_{1,1},\cdots,\theta_{1,N}$, the phases of $y_{1},\cdots,y_{N}$ cannot be controlled independently since the amplitudes of $y_{1},\cdots,y_{N}$ exhibit correlation.
The effect of this correlation is set aside for future research.
\end{remark}

\subsection{Comparison between systems with multiple cooperative RISs}
\label{Discussion on multi-RIS system}

\begin{figure}[!t]
\centering
	\includegraphics[trim = 40 270 450 500, clip, width=.9\linewidth]{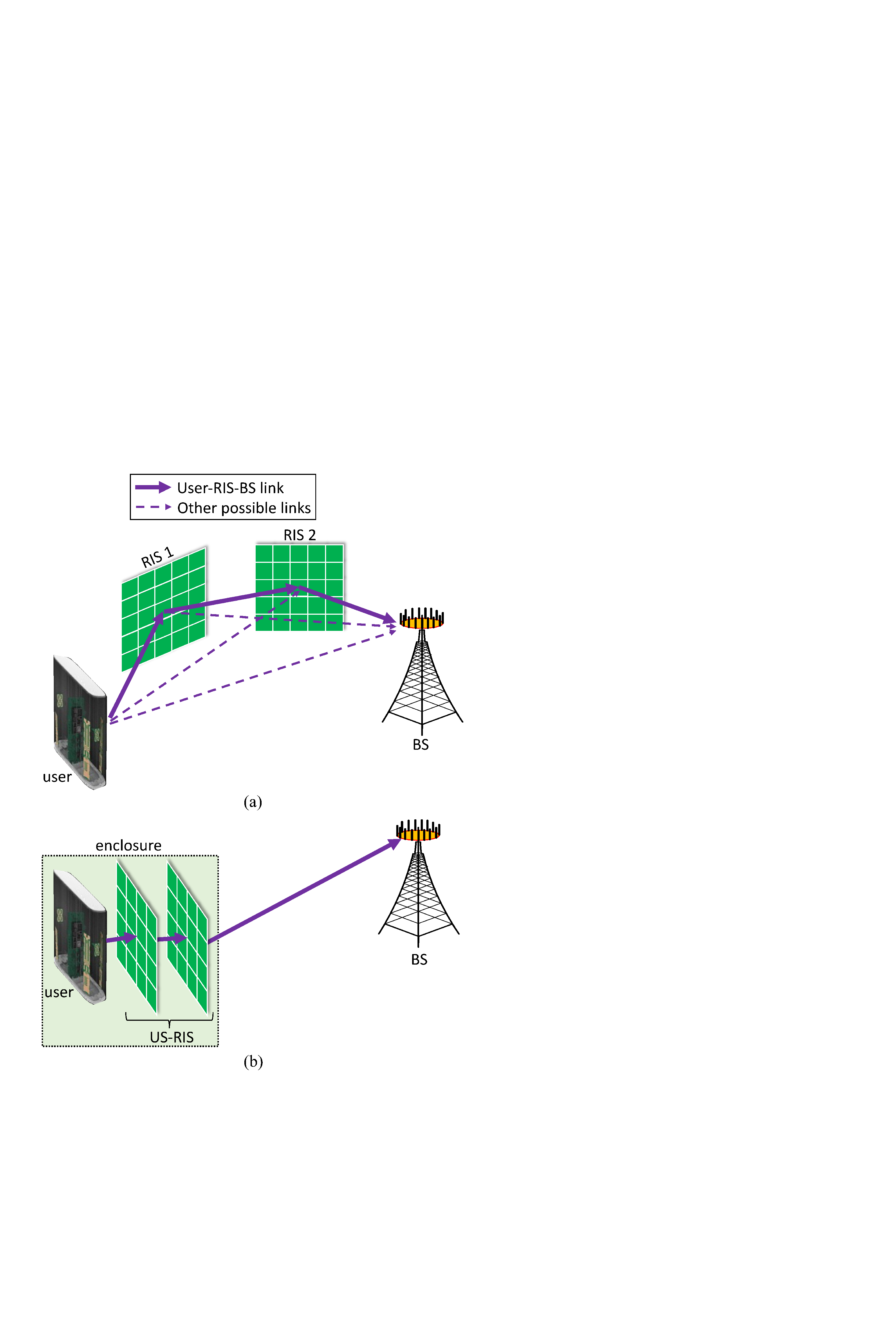}
	\caption{
	Different systems employing multiple cooperative RISs.
		(a) Multiple cooperative BSS-RIS-aided communications.
		(b) Multi-layer US-RIS-aided communications.
		}
	\label{compare}
\end{figure}

As revealed in the above theoretical analysis, when relying on collaborative phase shift control in our multi-layer US-RIS, the amplitude term of signals can be partially controlled.
Indeed, several other contributions on multiple cooperative BSS-RIS-aided communications have also done so \cite{zheng2021double,you2020wireless}.
As shown in Fig. \ref{compare} (a), iin multiple cooperative BSS-RIS-aided communications, multiple relay-like reflective BSS-RIS are employed for enhancing the system's performance, and the phase shift controls of different RISs are cooperatively optimized to achieve higher array gains.
In contrast to BSS-RIS-aided communications, the proposed multi-layer US-RIS-aided regime employs multiple transmissive RISs at the user side, as shown in Fig. \ref{compare} (b).

Indeed, both systems are capable of increasing array gains by cooperative beamformer design upon mitigating the “multiplicative fading” effect, which becomes aggravated by the multiple reflections/transmissions.
In multiple cooperative BSS-RIS-aided communications, one can increase the number of elements in relay-like BSS-RISs to compensate for the pathloss, but in multi-layer US-RIS-aided systems, the number of RIS elements is limited by the space constraint.
However, similar to the analysis in \cite{wu2021intelligent,you2021wireless}, the US-RIS are employed close to the user, which has the beneficial effect of reducing the multiplicative fading.

Moreover, since the multi-layer US-RIS is composed of transmissive surfaces, the energy loss when the signal penetrates the surface would be higher than that of the signal reflected from a surface.
However, the proposed multi-layer US-RIS-aided system has a valuable benefit and an associated difference in the system model, since the signal is only transmitted through the user-RIS-BS link thanks to the enclosure-based construction.
By contrast, the multiple cooperative BSS-RIS-aided system comprises many other possible links, including all the single-reflection links and the direct link.
Hence, additional channels have to be estimated, which requires more pilots.

\section{Multi-Layer Transmit Beamformer Design}
\label{Precoding-Design}
To further investigate the proposed US-RIS, in this section, we extend our discussions to the more general case of $L\geq 2$ and $K\geq 1$.
Based on the system model developed, we first formulate our \ac{SNR} maximization problem in Subsection~\ref{Problem formulation}, which is non-convex.
We will circumvent this challenge by developing a \ml \tb design relying on an iterative algorithm.
Then, the overview of our \ml transmit beamformer design is provided in Subsection~\ref{Overview of the proposed precoding design}, while the intricate details of the optimization process are discussed in Subsection~\ref{Optimal beamforming, US-RIS precoding, and combining}.
We also discuss the possible extension of the proposed method to the multi-user case in Subsection~\ref{Extension to multi-user case}.
Finally, the convergence and the computational complexity of the \ml \tb design are analyzed in Subsection~\ref{Convergence and complexity analysis}.

\subsection{Problem formulation}
\label{Problem formulation}
Based on the system model of Fig. \ref{MIMO} in Section~\ref{System Model}, we aim for maximizing the detection \ac{SNR} of the proposed US-RIS-aided system.
Observe from the uplink signal $z$ combined at the BS as shown in \eqref{CombinedSignal} that an equivalent noise contribution of $\v\H\bm n$ is introduced, which obeys $\v\H\bm n\sim\mathcal{CN}\left(0,\left\Vert \v\H\right\Vert_{2}^{2}\sigma^{2}\right)$.
Hence, the detection SNR at the BS can be represented as
\begin{equation}
\label{SNR}
\text{SNR}=\frac{\left\vert
\bm v\H\g \H \left( \prod_{l=L}^{1}\kappa\T_{l}\f_{l}\right)\w \right\vert^{2}
}{\left\Vert \v\H\right\Vert_{2}^{2}\sigma^{2}}.
\end{equation}
Then, given the maximum transmit power constraint of the user and phase shifts constraint at the US-RIS,
the SNR maximization problem is formulated as
\begin{subequations}
\label{problem}
\begin{align}
\label{objective}
\max_{\v,\T_{1},\cdots,\T_{L},\w}&\text{SNR}=\frac{\left\vert
\bm v\H\g \H \left( \prod_{l=L}^{1}\kappa\T_{l}\f_{l}\right)\w \right\vert^{2}
}{\left\Vert \v\H\right\Vert_{2}^{2}\sigma^{2}},\\
\label{C1}
~~~~~\text{s.t.~~~~~}&\text{C}_1:~\left\Vert\w\right\Vert_{2}^{2}\leq\Pmax,\\
\label{C2}
~~~~~~~~~~~~~&\text{C}_2:~\left\vert\theta_{l,n}\right\vert=1,~\forall l,n.
\end{align}
\end{subequations}

Due to the non-convexity of the objective function \eqref{objective} and the constraint \eqref{C2}, the joint optimization of both the \ac{RC} vector $\v$, as well as of the US-RIS \ac{TPS} matrices $\T_{1},\cdots,\T_{L}$, and of the \ac{UL-TBF} vector $\w$ is challenging.
To tackle this challenge, we propose an iterative algorithm as our \ml transmit beamformer design.

\subsection{Overview of the proposed \ml transmit beamformer design}
\label{Overview of the proposed precoding design}

The algorithm of solving this joint optimization problem \eqref{problem} is summarized in \textbf{Algorithm 1}.
Specifically, given the input channel matrices and parameters, the coupled variables $\v$, $\T_{1},\cdots,\T_{L}$, and $\w$ can be optimized by alternately updating one variable with the other variables fixed.
Once convergence of the objective function is reached, the iterations are curtailed, and the optimal beamformer design is found.
Generally, the computation of the \ml transmit beamformer design takes place at the BS side, and the BS should transmit the optimal UL-TBF vector and the TPS matrices back to the user.

\begin{algorithm}[!t]
\label{PrecodingDesign}
\caption{Multi-layer transmit beamformer design for US-RIS-aided communications}
\begin{algorithmic}[1]
\REQUIRE 
Channel matrices $\bm f_{1},\cdots,\bm f_{L}$, and $\bm g$;
maximum transmit power $\Pmax$.
\ENSURE 
Optimized \ac{RC} vector $\bm v$;
optimized US-RIS \ac{TPS} matrices $\T_{1},\cdots,\T_{L}$;
optimized \ac{UL-TBF} vector $\bm w$;
maximized SNR.

		\STATE Initialize $\v$, $\T_{1},\cdots,\T_{L}$, and $\w$;
		\WHILE {no convergence of SNR}
		\STATE Update $\bm v\opt$ by \eqref{v_optimal};
		\STATE Update $\bm \T_{1}\opt,\cdots,\bm \T_{L}\opt$ in turn by \eqref{T_optimal};
		\STATE Update $\bm w\opt$ by \eqref{w_optimal};
		\STATE Update SNR by \eqref{SNR};
		
		\ENDWHILE	
		\RETURN $\v$, $\T_{1},\cdots,\T_{L}$, $\w$, and SNR.
\end{algorithmic}
\end{algorithm}

\subsection{Optimal \ac{UL-TBF}, \ac{TPS}, and \ac{RC} of Fig. \ref{MIMO}}
\label{Optimal beamforming, US-RIS precoding, and combining}
In this subsection, we derive the closed-form expression 
of the optimal solution corresponding to each variable.
For ease of notation, we first define
\begin{equation}
\label{xi}
\bm \xi_{(p,q)}=
\left \{
	\begin{aligned}
	&\prod_{l=p}^{q}\kappa\T_{l}\f_{l},&&p\in[L], q\in[L],\\
	&\bm I_{N},&&p=L, q=L+1,\\
	&\bm I_{K},&&p=0, q=1.\\
	\end{aligned}
	\right.
\end{equation}
Then, the updates of different variables are respectively provided as follows.

\subsubsection{Optimal \ac{RC}}
For determining the \ac{RC} vector $\bm v$ of Fig. \ref{MIMO}, while fixing all the other variables, the optimization problem \eqref{problem} can be equivalently reformulated as
\begin{equation}
\label{v_reformulate}
\max _{\bm v}~\text{SNR}=\frac{
\bm v\H\g \H \x_{(L,1)}\w \w \H\x\H_{(L,1)}\g\v
}{\left\Vert \v\H\right\Vert_{2}^{2}\sigma^{2}}:=
\frac{
\bm v\H\bm U\v
}{\left\Vert \v\H\right\Vert_{2}^{2}\sigma^{2}},
\end{equation}
where $
\bm U=\g \H \x_{(L,1)}\w \w \H\x\H_{(L,1)}\g$
is a positive semidefinite matrix.
Based on matrix analysis,
the maximum SNR can be achieved when $\v$ is
the eigenvector of $\bm U$, corresponding to its largest
eigenvalue, which is formulated as
\begin{equation}
\label{v_optimal}
\v \opt= \bm \psi_{\text{max}}\left(
\g \H \x_{(L,1)}\w \w \H\x\H_{(L,1)}\g
\right).
\end{equation}

\subsubsection{Optimal US-RIS \ac{TPS}}
For the \ac{TPS} matrix $\T_{l}~(l\in [L])$,
since $\T_{l}\equiv\text{diag}(\t_{l})$ is a diagonal matrix
and $\f_{l}\x_{(l-1,1)}\w$ is a column vector,
we have
\begin{equation}
\label{tran}
\x_{(l,1)}\w=\text{diag}(\f_{l}\x_{(l-1,1)}\w)\t_{l}.
\end{equation}
By exploiting the transformation \eqref{tran}, the SNR can be rewritten as
\begin{equation}
\begin{aligned}
\text{SNR}&=\frac{\left\vert
\bm v\H\g \H \x_{(L,1)}\w \right\vert^{2}
}{\left\Vert \v\H\right\Vert_{2}^{2}\sigma^{2}}\\&=
\frac{\left\vert
\bm v\H\g \H \x_{(L,l+1)}\x_{(l,1)}\w \right\vert^{2}
}{\left\Vert \v\H\right\Vert_{2}^{2}\sigma^{2}}\\&=
\frac{\left\vert
\bm v\H\g \H \x_{(L,l+1)}\text{diag}(\f_{l}\x_{(l-1,1)}\w)\t_{l} \right\vert^{2}
}{\left\Vert \v\H\right\Vert_{2}^{2}\sigma^{2}}.
\end{aligned}
\end{equation}
Because of the constraint \eqref{C2}, the optimal value
of $\t_{l}$ is obtained when all entries of $\t_{l}$
have the complementary phase angle as the vector
multiplied on the left, which is expressed as
\begin{equation}
\label{T_optimal}
\t _{l}\opt =\text{exp}\left(j
\text{arg}\left(
\text{diag}
\left(
\f_{l}\x_{(l-1,1)}\w
\right)\H\x_{(L,l+1)}\H\g\v
\right)
\right),~\forall l\in[L].
\end{equation}

\subsubsection{Optimal \ac{UL-TBF}}
Finally, for the \ac{UL-TBF} vector $\w$, 
we first consider the optimization of the
normalized vector $\left\langle \w \right\rangle$, i.e.,
\begin{subequations}
\label{w_tran}
\begin{align}
\label{w_tranobjective}
&\max_{\left\langle \w \right\rangle}~\frac{\left\vert
\bm v\H\g \H \x_{(L,1)}\left\langle \w \right\rangle \right\vert^{2}
}{\left\Vert \v\H\right\Vert_{2}^{2}\sigma^{2}}=\frac{\text{SNR}}{\left\Vert\w\right\Vert_{2}^{2}},\\
\label{w_tranC1}
&\text{~s.t.~~~~}\text{C}_1:~\left\Vert\left\langle \w \right\rangle\right\Vert_{2}=1.
\end{align}
\end{subequations}
We then obtain the optimized $\left\langle \w \right\rangle$ as
\begin{equation}
\left\langle \w \right\rangle\opt=\left\langle
\x_{(L,1)}\H\g\v
\right\rangle.
\end{equation}
Upon considering the constraint \eqref{C1} and the relationship between the optimized $\w$ for the subproblem \eqref{problem} and the optimized $\left\langle \w \right\rangle$ for the subproblem \eqref{w_tran}, we arrive at
\begin{equation}
\label{w_optimal}
\w\opt=\sqrt{\Pmax}\left\langle \w \right\rangle\opt
=\sqrt{\Pmax}\left\langle
\x_{(L,1)}\H\g\v
\right\rangle.
\end{equation}

\subsection{Extension to multi-user case}
\label{Extension to multi-user case}
In Subsection~\ref{Overview of the proposed precoding design} and \ref{Optimal beamforming, US-RIS precoding, and combining}, we have proposed a \ml transmit beamformer design for US-RIS-aided communications that supports a single multi-antenna-aided user.
For the multi-user case, we can simply extend the analysis relying on signal model \eqref{CombinedSignal}.
Explicitly, we may assume that $U$ users are supported by an $L$-layer US-RIS having $N$ elements on each layer.
Upon extending our notations in Subsection~\ref{US-RIS-aided communications}, let us assume that $\bm w_{u}, \T_{u,1}, \cdots, \T_{u,L}$ denote the independent UL-TBF vector and TPS matrices for the $u$-th user, respectively.
Then, the signal combined at the BS receiver can be rewritten as
\begin{equation}
\label{CombinedSignal_MultiUser}
z=\v\H\sum_{u=1}^{U}\g_{u} \H \left( \prod_{l=L}^{1}\kappa\T_{u,l}\f_{u,l}\right)\w_{u} s+\v\H\bm n,
\end{equation}
where $\bm g_{u}$ and $\bm f_{u,l}$ denote the channels corresponding to the $u$-th user spanning from the US-RIS's $L$-th layer to the BS and from the user to the US-RIS's first layer, respectively.
Therefore, the new objective of the joint optimization, which now becomes the sum-rate of all users, can be represented as
\begin{equation}
\label{sum-rate}
R_{\text{sum}}=\sum_{u=1}^{U}\log_{2}\left(1+\text{SINR}_{u}\right),
\end{equation}
where the received \ac{SINR} of the $u$-th user at the BS can be represented as
\begin{equation}
\label{SINR}
\begin{aligned}
&\text{SINR}_{u}\\
&=\frac{\left|\bm v^{H}\bm g_{u}^{H}\left(\prod_{l=L}^{1}\kappa \T_{u,l}\bm f_{u,l}\right)\bm w_{u}\right|^{2}}{\sum_{u'\neq u}\left|\bm v^{H}\bm g_{u'}^{H}\left(\prod_{l=L}^{1}\kappa \T_{u',l}\bm f_{u',l}\right)\bm w_{u'}\right|^{2}+\left\Vert\bm v ^{H}\right\Vert_{2}^{2}\sigma^{2}}.
\end{aligned}
\end{equation}
Following the same constraints of the maximum transmit power and the phase shifts as for the US-RIS, the new optimization problem can be solved by our proposed \ml transmit beamformer design of the single-user case by employing standard fractional programming methods to tackle the non-convex nature of the objective.
However, the full characterization of this multi-user scenario requires a full journal paper to be conceived in our future research.

\subsection{Algorithm analysis}
\label{Convergence and complexity analysis}
\subsubsection{Convergence analysis}
\label{convergence analysis}
For the proposed \ml transmit beamformer design, we adopt the general
assumption that only the phase
shifts of the elements can be continuously controlled.
Under this assumption, 
if the variables are alternately updated according to
\eqref{v_optimal}, \eqref{T_optimal}, and \eqref{w_optimal}, which is the optimal solution of the corresponding subproblem with all the other variables fixed, the objective function, namely the SNR, will increase monotonically, hence indicating the strict convergence of the \ml transmit beamformer design.

\subsubsection{Complexity analysis}
\label{Complexity analysis}
The overall computational complexity of the proposed \ml transmit beamformer design is dominated by the updates of the variables $\v$, $\T_{1},\cdots,\T_{L}$, and $\w$, as shown in \eqref{v_optimal}, \eqref{T_optimal}, and \eqref{w_optimal}, respectively. 
Note however that, the $\x_{(p,q)}$ terms associated with different $p$ and $q$ have strong correlation that may be exploited for reducing the redundancy of the repeated matrix multiplications.
Specifically, all $\x_{(p,q)}$ terms involved in \eqref{v_optimal}, \eqref{T_optimal}, and \eqref{w_optimal} can be obtained throughout the computational processes of $\x_{(L,1)}$ and $\x_{(L,2)}$, hence only the computational complexities of $\x_{(L,1)}$ and $\x_{(L,2)}$ actually count in the $\x_{(p,q)}$ terms of the parameter updates.

\begin{table}[t] \vspace{0.6em}
\caption{Complexity of updating each variable}
\label{ComplexityTable}
\centering
\begin{tabular}{|l|r|}
    \hline
    \textbf{Variable}&\textbf{Complexity} $\mathcal O(\cdot)$\\\hline
$\x_{(L,1)}$&$LNK+(L-1)N^{2}K$\\\hline
$\x_{(L,2)}$&$(L-1)N^{2}+(L-2)N^{3}$\\\hline
$\v\opt$&$NK+MN+M^{2}$\\\hline
$\T_{1}\opt,\cdots,\T_{L}\opt$~~~~&~~~~~~$L\left(NK+N^{2}K+MN+N^{2}\right)$\\\hline
$\w\opt$&$MN+MNK+K$\\\hline
    \end{tabular}
\end{table}

We have summarized the computational complexities of the variables
in TABLE \ref{ComplexityTable}.
Since the number of RIS elements
is usually high \cite{wang2020intelligent},
we can reasonably assume that
$N\gg K$ and $N^{2}\gg M$.
Furthermore, due to the loss factor characterizing the process when the electromagnetic waves penetrate each layer, the number of layers $L$ must not be large compared to $N$, i.e. we have $N\gg L$.
Under these assumptions, the overall complexity of the proposed \ml precoding design in a single iteration may be approximated by ${\mathcal O}\left(N^{3}+MNK+M^{2}\right)$.
For comparison, the complexity analysis of the beamformer design in BSS-RIS-aided communications can be regarded as a special case of the US-RIS-aided scenario, where the BSS-RIS has a single-layer structure associated with $LN$ elements, which is the same as the total number of elements in our US-RIS.
Therefore, the complexity of a single iteration in the \ml \tb design of BSS-RIS-aided communications is ${\mathcal O}\left(N^{3}L^{3}+MNKL+M^{2}\right)$.
Since we have $N\gg L$, the computational complexity of the proposed \ml \tb design is almost the same as that of the traditional BSS-RIS-aided scheme, which is affordable in practice.

\section{Simulation Results}
\label{Simulation Results}
In this section, we provide extensive simulation results for quantifying the performance of the proposed US-RIS concept and of the corresponding \ml \tb design for US-RIS-aided communications.
The simulation scenario is introduced first in Subsection~\ref{Simulation setup}.
Then, numerical results are discussed in the following subsections.

\subsection{Simulation setup}
\label{Simulation setup}

\begin{figure}[!t] 
   \centering
   \includegraphics[trim = 40 30 190 90, clip, width=.95\linewidth]{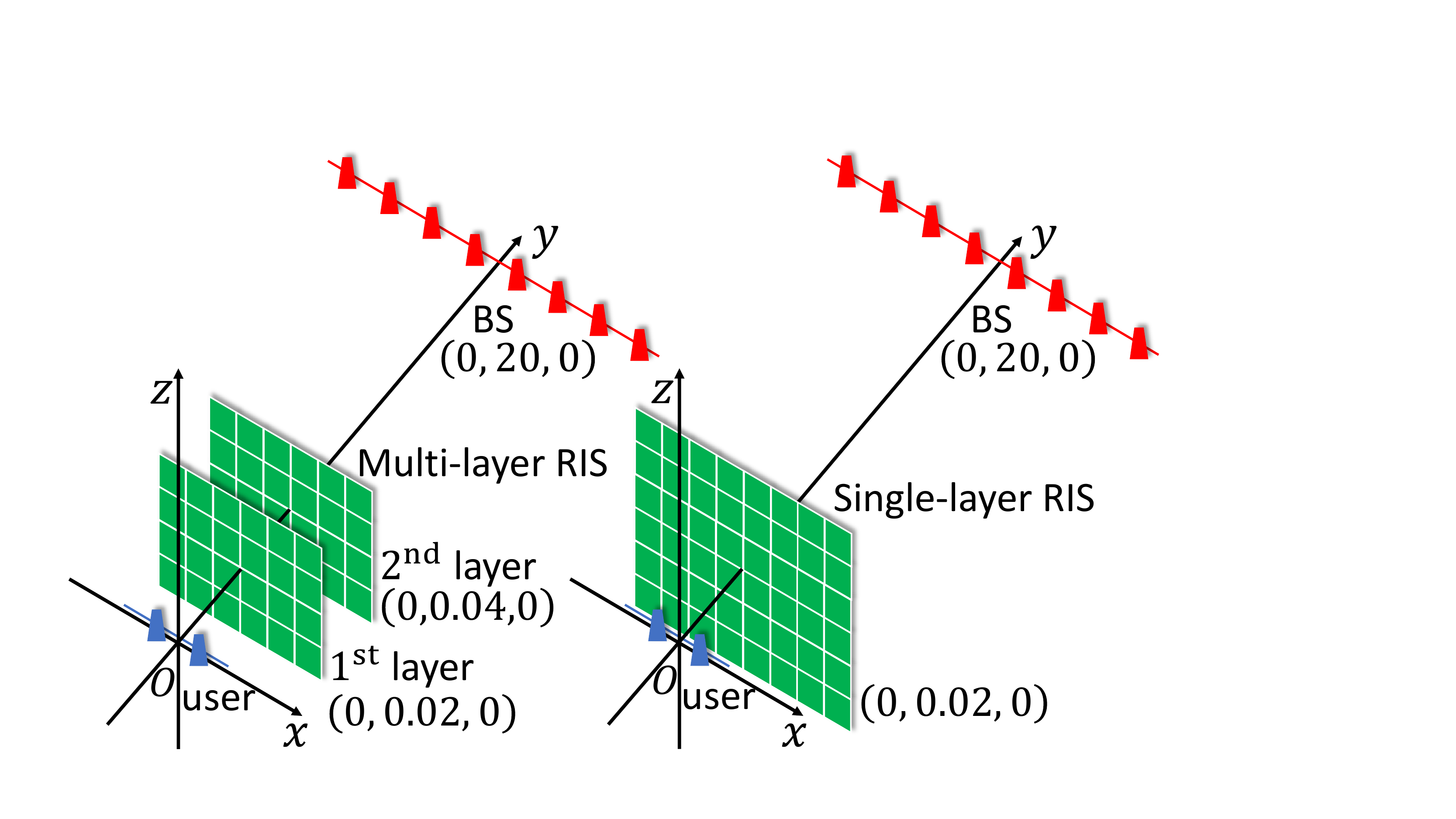} 
    \centerline{(a)~~
    ~~~~~~~~~~~~~~~~~~~~~~~~~~~~~(b)}
   \caption{The simulation scenario where
   different types of RISs are employed to assist communications.
   (a) Multi-layer US-RIS-aided communications.
   (b) Single-layer US/BSS-RIS-aided communications.}
   \label{setup}
\end{figure}

For simplicity but without loss of generality, we consider a 3-D scenario where different types of RISs are employed for uplink transmission from the user to the BS.
Specifically, the topology considered is shown in Fig. \ref{setup}. 
Let $\lambda$ denote the wavelength of the transmitted uplink signal.
We assume that both the user and the BS are equipped with a uniform linear array (ULA) having $2$ and $8$ antennas, respectively, where the distance between the adjacent elements is $\frac{\lambda}{2}$.
A multi-layer US-RIS, with $2$ layers and $8\times12$ elements on each layer,
is used for enhancing the user's uplink transmission, as shown in Fig. \ref{setup} (a). 
For comparison, we consider another two cases, where a single-layer US-RIS and a single-layer reflective BSS-RIS having $12\times16$ elements are respectively used at the same position as the multi-layer US-RIS's first layer shown in Fig. \ref{setup} (b). 
The total number of elements in the multi-layer US-RIS, single-layer US-RIS, and single-layer BSS-RIS are the same, and they are all uniform planar arrays (UPAs), where the elements of size $\frac{\lambda}{2}\times \frac{\lambda}{2}$ are closely packed with no spacing \cite{bjornson2019massive}.
We assume that the user, the multi-layer US-RIS's first layer (located at the same position as the single-layer US/BSS-RIS),
the multi-layer US-RIS's second layer, and the BS are located with their centers at $(0\,\text{m},0\,\text{m},0\,\text{m})$, $(0\,\text{m},0.02\,\text{m},0\,\text{m})$, $(0\,\text{m},0.04\,\text{m},0\,\text{m})$, and $(0\,\text{m},20\,\text{m},0\,\text{m})$, respectively.
The frequency of the transmitted signal is set to $f=2.5\,\text{GHz}$. The noise power is set to $\sigma^{2}=1\times 10^{-6}$. The loss factor $\kappa$ when a wave penetrates a transmissive RIS is set to $0.8$.

As for the channel model, we categorize all channels involved into two types.
The RIS-BS channel is the Type \RNum{1} channel which is a far-field channel assumed to obey the free-space propagation, where an ideal transmit antenna sends a signal to a lossless receive antenna.
The pathloss of the Type \RNum{1} channel can be directly obtained from Friis' formula.
On the other hand, the user-RIS and RIS-RIS channels are termed as the Type \RNum{2} channels.
To elaborate, the Type \RNum{2} channel is the near-field channel spanning from a lossless antenna to an element of the UPA under the \ac{LoS} assumption.
The lossless transmit antenna can be either an element of a ULA or an element of a UPA, corresponding to the user-RIS and RIS-RIS channels.
For the pathloss of the Type \RNum{2} channel, we adopt the exact expression introduced in \cite[Lemma 1]{bjornson2020power}, which calculates the free space channel gain from a lossless antenna to a planar array.
Note that, we only consider the user-RIS-BS links, and the other potential links are ignored, as mentioned in Section~\ref{System Model}.

Finally, we assume for the proposed \ml \tb design that the channel state information (CSI) is perfectly known \cite{guo2019weighted,Zhang'20}, which can be estimated through RIS-based channel estimation and channel feedback methods in practice \cite{Huchen,an2020optimal,ma2021wideband}.
Specifically, in US-RIS-aided communications, the channel spanning from the user to the US-RIS and that between adjacent layers of the US-RIS can be estimated beforehand, thanks to stability ensured by the enclosure.
As for the initializations, all RISs are initialized by random phase values in the feasible set, while the \ac{UL-TBF} and \ac{RC} vectors are initialized as ones in our iterative algorithm.

\subsection{Performance of the US-RIS-aided communications}
\label{Performance of received signal under US-RIS aided system}

\begin{figure}[!t] 
   \centering
   \includegraphics[trim = 30 0 30 20, clip, width=.9\linewidth]{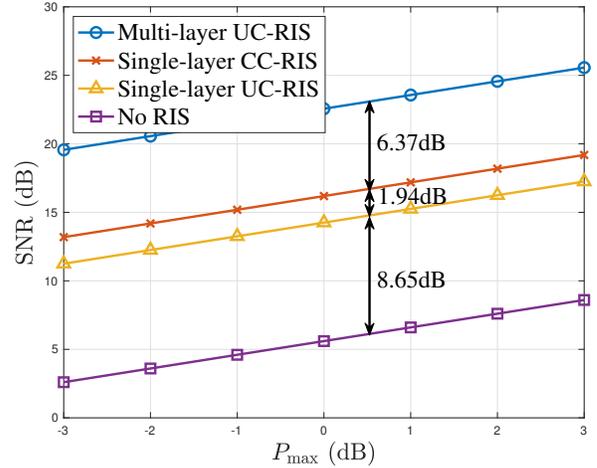} 
   \caption{Detection SNR versus the maximum transmit power $P_{\text{max}}$.}
   \label{P2SNR}
\end{figure}

In Fig. \ref{P2SNR}, we portray the detection SNR versus the maximum transmit power $P_{\text{max}}$ for different RIS scenarios.
We also consider a \textit{“No RIS”} benchmarker for comparison \cite{Zhang'20}, where the RIS shown in Fig. \ref{setup} is removed but the optimization of \ac{UL-TBF} vector and \ac{RC} vector based on the Type \RNum{1} UE-BS channel is retained.

Observe from Fig. \ref{P2SNR} that the SNR versus the maximum transmit power $P_{\text{max}}$ exhibits a near-linear relationship, which can also be observed from the parameter updates \eqref{v_optimal}, \eqref{T_optimal}, and \eqref{w_optimal}. 
Therefore, we can readily compare different scenarios with the power constraint $P_{\text{max}}$ fixed.
The associated differences are marked in Fig. \ref{P2SNR}.
Observe that the communication without RISs has the lowest SNR, illustrating that all three RIS scenarios can substantially improve the uplink transmission. 
Another observation is that the single-layer US-RIS-aided communications has a $1.94\,\text{dB}$ loss compared to its single-layer BSS-RIS-aided counterpart, which is caused by the loss when a wave penetrates the RIS. 
Moreover, the multi-layer US-RIS-aided system achieves an increase of $8.31\,\text{dB}$ detection SNR over its single-layer US-RIS-aided counterpart.
Note that in this content we assume both US-RIS structures to have the same total number of RIS elements, so the DoF in phase exploited by our transmit beamformer design is the same.
Thus, logically, the increment of detection SNR arises from the additional DoF in amplitude provided by the multi-layer structure, which is an explicit benefit over the conventional single-layer structure that we have shown theoretically.

\begin{figure}[!t] 
   \centering
   \includegraphics[trim = 30 0 30 20, clip, width=.9\linewidth]{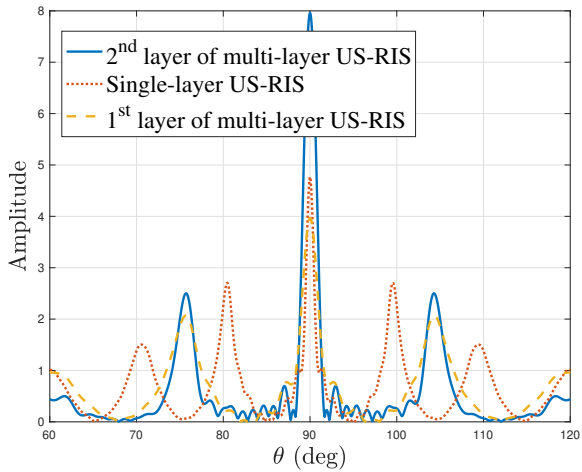} 
   \caption{Radiation pattern of different RISs.}
   \label{Pattern}
\end{figure}


For further characterizing the \ml structure, in Fig. \ref{Pattern} we present the radiation pattern of different RISs and compare the quality of the beam by comparing their mainlobes and sidelobes.
Naturally, a beam having lower sidelobes and a higher mainlobe is preferred.
Observe from Fig.~\ref{Pattern} that the transmitted beam of the first layer in the multi-layer US-RIS has a lower mainlobe than the single-layer US-RIS, which is caused by the reduced number of elements in the first layer of the multi-layer US-RIS.
However, after penetrating the second layer and obtaining a new DoF in amplitude terms, the mainlobe to sidelobe ratio is considerably improved.
The above results coincide well with our theoretical analysis provided in Section~\ref{Performance Analysis}.

\subsection{Convergence of the proposed \ml \tb design}
\label{Convergence of proposed precoding design}

\begin{figure}[!t] 
   \centering
   \includegraphics[trim = 30 0 30 20, clip, width=.9\linewidth]{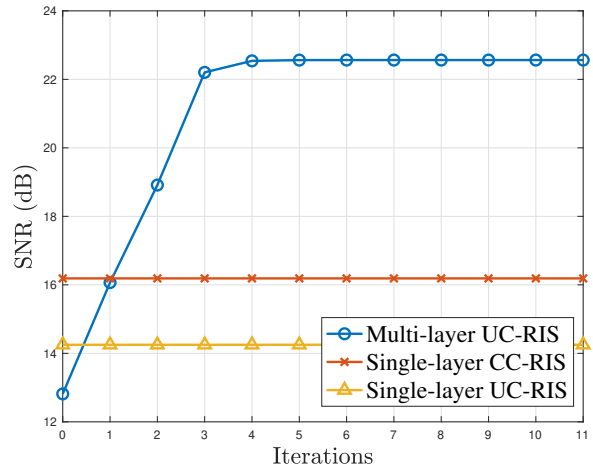} 
   \caption{Detection SNR at $P_{\text{max}}=0\text{dBW}$ against the number of iterations.}
   \label{conv}
\end{figure}

Recall from Subsection~\ref{convergence analysis} that, the proposed \ml \tb design exhibits monotonic convergence.
In Fig. \ref{conv}, we further characterize its convergence in the simulation scenario considered. 
The results of Fig. \ref{conv} clearly illustrate that, the proposed \ml \tb design for multi-layer US-RIS converges within $4$ to $5$ iterations, while the conventional single-layer BSS-RIS-aided scheme converges within $1$ to $2$ iterations.
However, our solution outperforms the latter after the first iteration.

\subsection{EAR Analysis on RIS}
\label{Analysis on RIS's element activation ratio}

\begin{figure*}[!t]
\begin{minipage}{0.273\linewidth}
  \centerline{\includegraphics[trim = 50 40 40 35, clip, width=.92\linewidth]{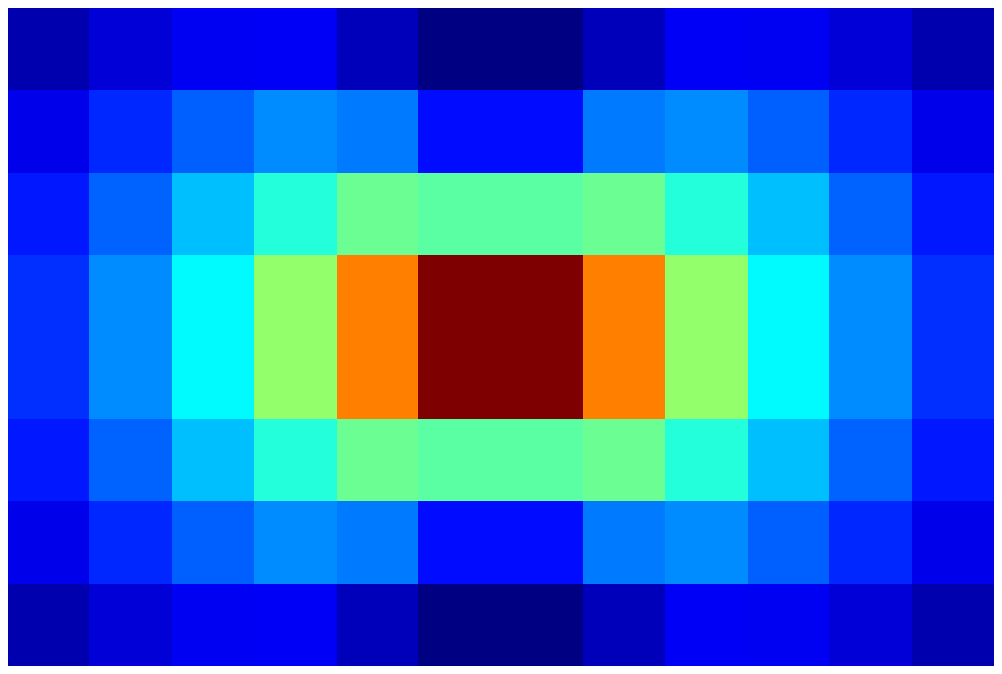}}
\end{minipage}
\hfill
\begin{minipage}{0.273\linewidth}
  \centerline{\includegraphics[trim = 50 40 40 35, clip, width=.92\linewidth]{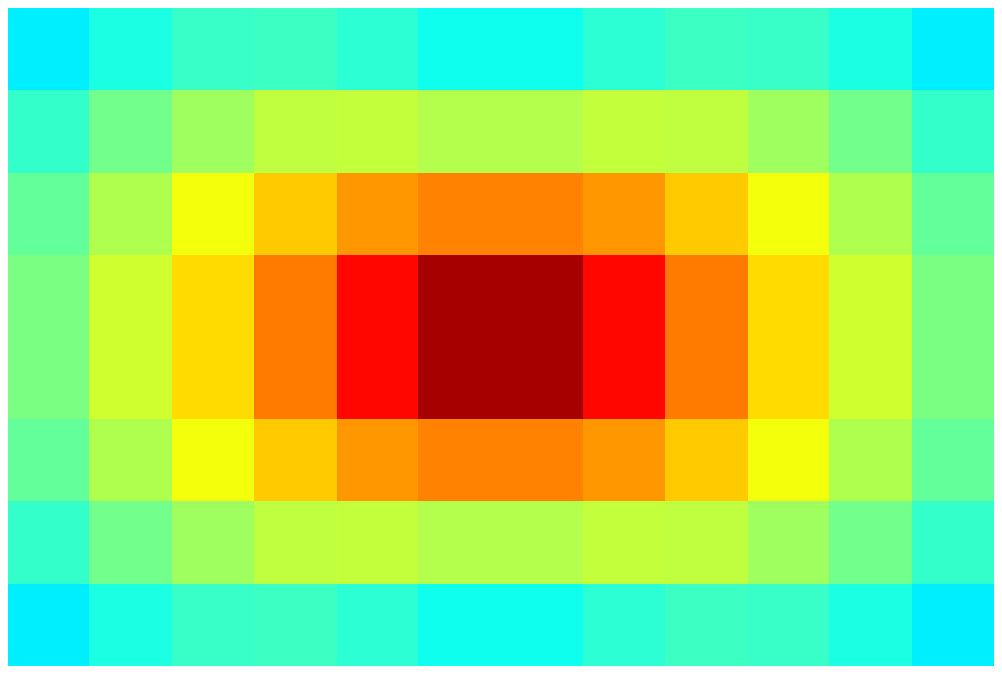}}
  \end{minipage}
  \hfill
\begin{minipage}{0.414\linewidth}
  \centerline{\includegraphics[trim = 65 50 35 35, clip, width=.9\linewidth]{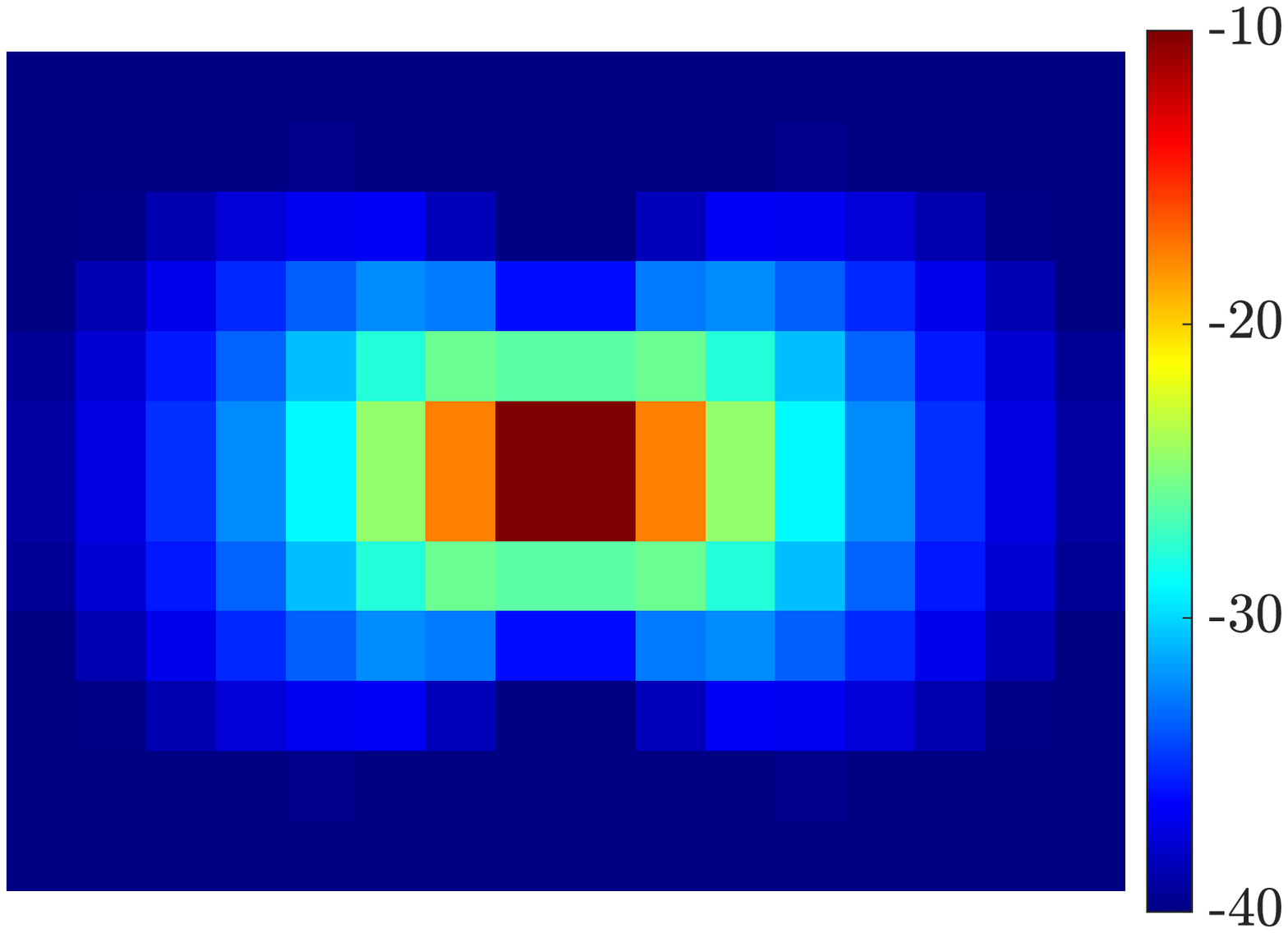}}
\end{minipage}
\centerline{(a)~~~~~~~~~~~~~~~~~~~~~~~~~~~~~~~~~~~~~~~(b)~~~~~~~~~~~~~~~~~~~~~~~~~~~~~~~~~~~~~~~~~~~~~~~~(c)~~~~~~~~~~~~~}
\caption{Power distribution of \ml and \sl US-RIS
(colorbar on the right is unified for 3 subfigures, unit: dBW),
implying different EARs of different RISs.
Threshold percentage is set to $\varepsilon=1/6$.
(a) The first layer of \ml US-RIS with EAR $29.2\%$.
(b) The second layer of \ml US-RIS with EAR $87.5\%$.
(c) Single-layer US-RIS with EAR $22.9\%$.}
\label{PowerDistribution}
\end{figure*}

To further reveal the benefits of the multi-layer structure in controlling the amplitude of the signal, the power distribution of both \ml and \sl US-RIS are presented in Fig. \ref{PowerDistribution}. 
Observe from Fig. \ref{PowerDistribution} (a) and (c) that both in the first layer of the \ml US-RIS and in the \sl US-RIS, the central elements are assigned most of the power, while the non-central elements have a limited contribution.
By contrast, as shown in Fig. \ref{PowerDistribution} (b), the power distribution is 
transformed to a more balanced power pattern in the second layer of US-RIS with the aid of the phase shift control of the first layer.
Although the power of the central elements is reduced, the elements along the edges become more influential.

To quantify this phenomenon, 
we apply the EAR metric defined in Subsection~\ref{Architecture}.
Explicitly, the EARs of different RISs with a threshold percentage $\varepsilon=1/6$ are shown in Fig. \ref{PowerDistribution}.
The first layer of the \ml US-RIS has an EAR of $29.2\%$, which is higher than that of the \sl US-RIS having an EAR of $22.9\%$, because the first layer of the \ml US-RIS has fewer elements.
Furthermore, as a benefit of the multi-layer structure, the EAR of the second layer of the US-RIS is $87.5\%$, and the overall EAR of the \ml US-RIS is $58.3\%$, which is twice higher than that of the \sl US-RIS.
The obvious improvement in EAR also demonstrates the ascendancy of the proposed architecture with \ml structure.

\section{Summary and Conclusions}
\label{Conclusions}
In this paper, we have proposed the concept of US-RIS
to circumvent the space-limit of employing a large-scale array at the user side.
In contrast to the existing BSS-RISs,
we have proposed a solution that is suitable for the user side.
Based on this concept, we have further proposed a novel architecture with the aid of a \ml US-RIS.
Our theoretical analysis has demonstrated that the amplitude of the signal penetrating the multi-layer US-RIS can also be partially controlled, which equips us with a new DoF. 
Furthermore, we have formulated the associated SNR maximization problem in the US-RIS-aided communications and have proposed the corresponding \ml \tb design that alternately finds the optimal solution for each variable.
Our simulation results have verified
the superiority of the proposed multi-layer US-RIS
as a realization of a large-scale array at the user side for uplink transmission.

However, a whole spate of open problems associated with \ml US-RIS requires further investigations in our future work. 
For example,
the theoretical analysis of this appealing multi-layer structure
is based on the assumption that the RIS phase shifts can be controlled continuously, while in practice only discrete phase shifts can be realized \cite{gong2020beamforming}. 
Furthermore, the RIS-element correlation has been neglected in this study, which requires special attention.
Finally, for practical hardware implementation considerations, the number of RIS elements in each layer is limited by the user's dimensions, and the design of transmissive RIS should reduce the energy loss and prohibit possible signal reflection.

\appendices
\section{Proof of Lemma \ref{lem}}
\label{Method}
Following the notations in Section~\ref{Performance Analysis},
the proof proceeds in two steps.
In the first step,
we derive the closed-form expression of $y_{n}$
by integration.
In the second step, under the assumption of controllable US-RIS phase shift matrices,
we analyze the range of phase and amplitude of $y_{n}$, respectively.

\subsection{Closed-form expression of $y_{n}$}
Denote the channel spanning from the point source $\bm s_{t}$ at the user to the arbitrary receiver point $\bm p=\left(p_{x},d_{1},p_{z}\right)$ on the US-RIS's first layer and the channel spanning from $\bm p$ to the center of the $(2,n)$-th element by $h_{1}\left(\bm s_{t},\bm p\right)$ and $h_{2}\left(\bm p,\bm s_{r}\right)$, respectively.
The expression of $h_{1}\left(\bm s_{t},\bm p\right)$
and $h_{2}\left(\bm p,\bm s_{r}\right)$ can be written as
\begin{subequations}
\label{ExactExpression}
	\begin{align}
	\label{ExactExpression1}
h_{1}\left(\bm s_{t},\bm p\right)&=
\left\vert
h_{1}\left(\bm s_{t},\bm p\right)
\right\vert
\exp\left(-j\frac{2\pi}{\lambda}
\left\Vert
\bm p-\bm s_{t}
\right\Vert
\right),
\\
\label{ExactExpression2}
h_{2}\left(\bm p,\bm s_{r}\right)&=
\left\vert
h_{2}\left(\bm p,\bm s_{r}\right)
\right\vert
\exp\left(-j\frac{2\pi}{\lambda}
\left\Vert
\bm s_{r}-\bm p
\right\Vert
\right),
	\end{align}
\end{subequations}
where $\lambda$ denotes the wavelength and
\begin{subequations}
\label{ChannelGain}
\begin{align}
\label{ChannelGain1}
\left\vert
h_{1}\left(\bm s_{t},\bm p\right)
\right\vert^{2}&=\frac{1}{4\pi}
\frac{d_{1}\left(p_{x}^{2}+d_{1}^{2}\right)}{
\left(p_{x}^{2}+p_{z}^{2}+d_{1}^{2}\right)^{5/2}},
\\
\label{ChannelGain2}
\left\vert
h_{2}\left(\bm p,\bm s_{r}\right)
\right\vert^{2}&=\frac{1}{4\pi}
\frac{d_{2}
\left[\left(p_{x}-\alpha_{n}\right)^{2}+d_{2}^{2}\right]}{
\left[\left(p_{x}-\alpha_{n}\right)^{2}+\left(p_{z}-\beta_{n}\right)^{2}+d_{2}^{2}\right]^{5/2}},
\end{align}
\end{subequations}
represent the channel gain along the $y$-direction, respectively
\cite{dardari2020communicating,bjornson2020power}.
Upon integrating over all receive points on the US-RIS's first layer
and considering the phase shift $\theta_{2,n}$
of the $(2,n)$-th element,
we arrive at the closed-form expression of
\begin{equation}
\label{ClosedForm}
y_{n}=\theta_{2,n}\sum_{j=1}^{N}
\iint_{\Omega_{1,j}}h_{1}\left(\bm s_{t},\bm p\right)
\theta_{1,j}
h_{2}\left(\bm p,\bm s_{r}\right)\text{d}\bm p.
\end{equation}

\subsection{Range of phase and amplitude}
As for the range of the phase $\arg(y_{n})$,
recall from \eqref{ClosedForm} that,
since the phase of $\theta_{2,n}$
can be controlled continuously in $[-\pi,\pi]$, 
the phase of $y_{n}$ can also
be controlled continuously in $[-\pi,\pi]$.

As for the range of amplitudes $\left\vert y_{n}\right\vert$,
we discuss the control strategy of RIS
required for achieving the minimum of $0$ and 
the maximum of $\zeta_{n}$, respectively.
Note that the phase shift matrix $\T_{2}$
does not influence the amplitude of $y_{n}$,
hence we can focus our attention on the phase shift matrix $\T_{1}$.

To obtain the maximum amplitude,
we have from \eqref{ClosedForm} that
\begin{equation}
\label{Cauchy}
\begin{aligned}
\left\vert y_{n}\right\vert^{2}&
\leq\sum_{j=1}^{N}
\iint_{\Omega_{1,j}}\left\vert h_{1}\left(\bm s_{t},\bm p\right)
h_{2}\left(\bm p,\bm s_{r}\right)\right\vert^{2}\text{d}\bm p\\&=\int_{-\frac{ab}{2}}^{\frac{ab}{2}}
\int_{-\frac{ab}{2}}^{\frac{ab}{2}}
\left\vert h_{1}\left(\bm s_{t},\bm p\right)\right\vert^{2}
\left\vert h_{2}\left(\bm p,\bm s_{r}\right)\right\vert^{2}
\text{d}p_{x}
\text{d}p_{z}.
\end{aligned}
\end{equation}
Upon substituting \eqref{ExactExpression}, 
we obtain \eqref{zeta_n}.

On the other hand, to obtain the minimum amplitude,
we construct a phase matrix $\T_{1}$ that results in $y_{n}=0$.
A potential construction method is constituted by
the following.
We define the four symmetric elements centered at
\begin{equation*}
\begin{aligned}
\bm s_{r,1}=\left(\alpha_{n},d_{1}+d_{2},\beta_{n}\right),&\quad
\bm s_{r,2}=\left(-\alpha_{n},d_{1}+d_{2},\beta_{n}\right),\\
\bm s_{r,3}=\left(\alpha_{n},d_{1}+d_{2},-\beta_{n}\right),&\quad
\bm s_{r,4}=\left(-\alpha_{n},d_{1}+d_{2},-\beta_{n}\right),
\end{aligned}
\end{equation*}
as a quaternion $\mathcal S_{n}=\left(\bm s_{r,1},\bm s_{r,2},
\bm s_{r,3},\bm s_{r,4}\right)$.
The $N$ elements on the first layer
can then be divided into $\frac{N}{4}$
quaternions that do not have intersection with each other.
We will show that, for each
quaternion $\mathcal S_{n}$,
there exist $\theta_{1},\theta_{2},\theta_{3},\theta_{4}$ values
which satisfy
\begin{equation}
\label{quaternion}
\sum_{j=1}^{4}
\theta_{j}
\iint_{\Omega_{1,j}}h_{1}\left(\bm s_{t},\bm p\right)
h_{2}\left(\bm p,\bm s_{r,j}\right)\text{d}\bm p=0.
\end{equation}
Then, 
for each quaternion,
we can construct a group of
phase shifts $\theta_{1},\theta_{2},\theta_{3},\theta_{4}$,
which has no contribution
to $y_{n}$ in \eqref{ClosedForm}.
With the aid of this process, we obtain a solution of $\T_{1}$ that
satisfies $y_{n}=0$.

In fact, based on the geometric properties,
the amplitudes of the four double integral terms in \eqref{quaternion} may form a
closed quadrilateral.
Hence there exist $\theta_{j}~(j\in[4])$,
with $\left\vert\theta_{j}\right\vert=1$,
that satisfy \eqref{quaternion}.

\section*{Acknowledgments}
We would like to thank Richard MacKenzie from BT Group plc, Mo Hao from Tsinghua SEM Advanced ICT LAB, and Dr. Chenghong Zhuang for their helpful discussion and support of this work.

\footnotesize
\balance 
\bibliographystyle{IEEEtran}
\bibliography{US-RIS, IEEEabrv}
\begin{IEEEbiography}[{\includegraphics[trim = 126 315 332 355, clip, width=1in,height=1.25in,keepaspectratio]{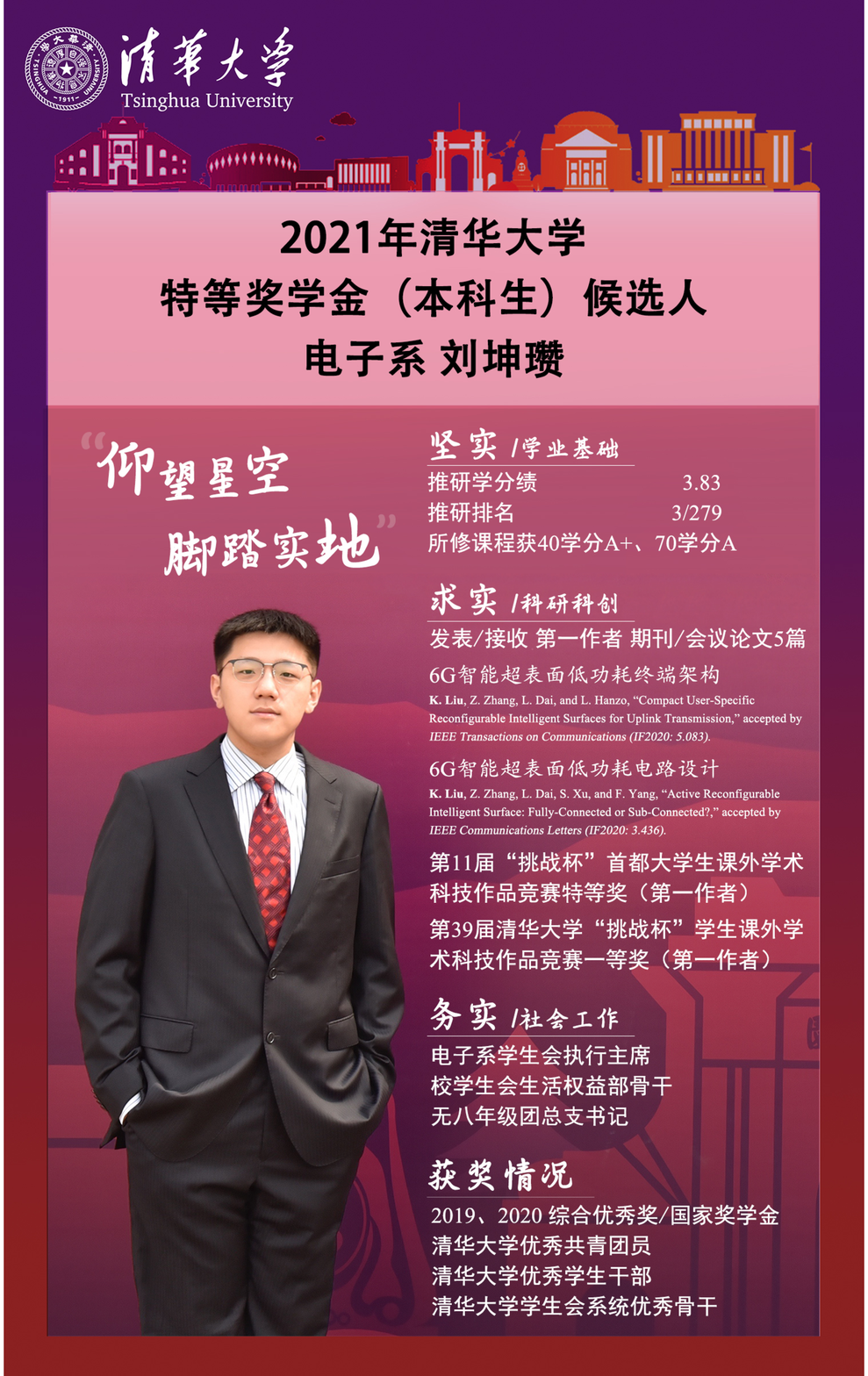}}]{Kunzan Liu}
has been pursuing the B.E. degree with the Department of Electronic Engineering, Tsinghua University, Beijing, China, since 2018. He joined the Tsinghua National Laboratory for Information Science and Technology (TNList) as a research student in 2020. His research interests include reconfigurable intelligent surface (RIS) and massive MIMO.
\end{IEEEbiography}

\begin{IEEEbiography}[{\includegraphics[trim = 0 60 0 5,width=1in,height=1.25in,clip,keepaspectratio]{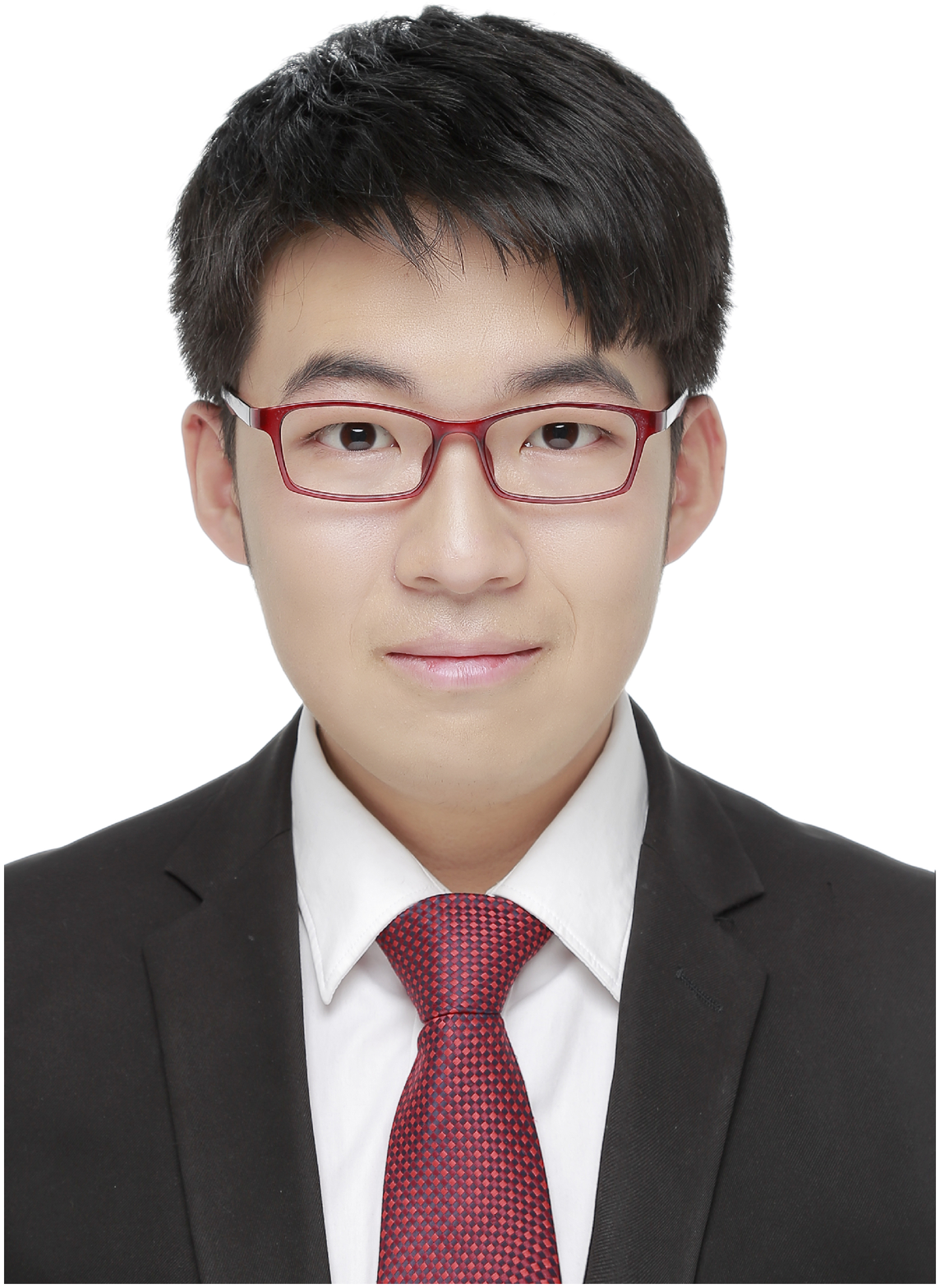}}]{Zijian Zhang}
(S'20) received the B.E. degree in electronic engineering from Tsinghua University, Beijing, China, in 2020. He is currently working toward the Ph.D. degree in electronic engineering from Tsinghua University, Beijing, China.
His research interests include physical-layer algorithms for massive MIMO and reconfigurable intelligent surfaces (RIS). He has received the National Scholarship in 2019.
\end{IEEEbiography}

\begin{IEEEbiography}[{\includegraphics[width=1in,height=1.25in,clip,keepaspectratio]{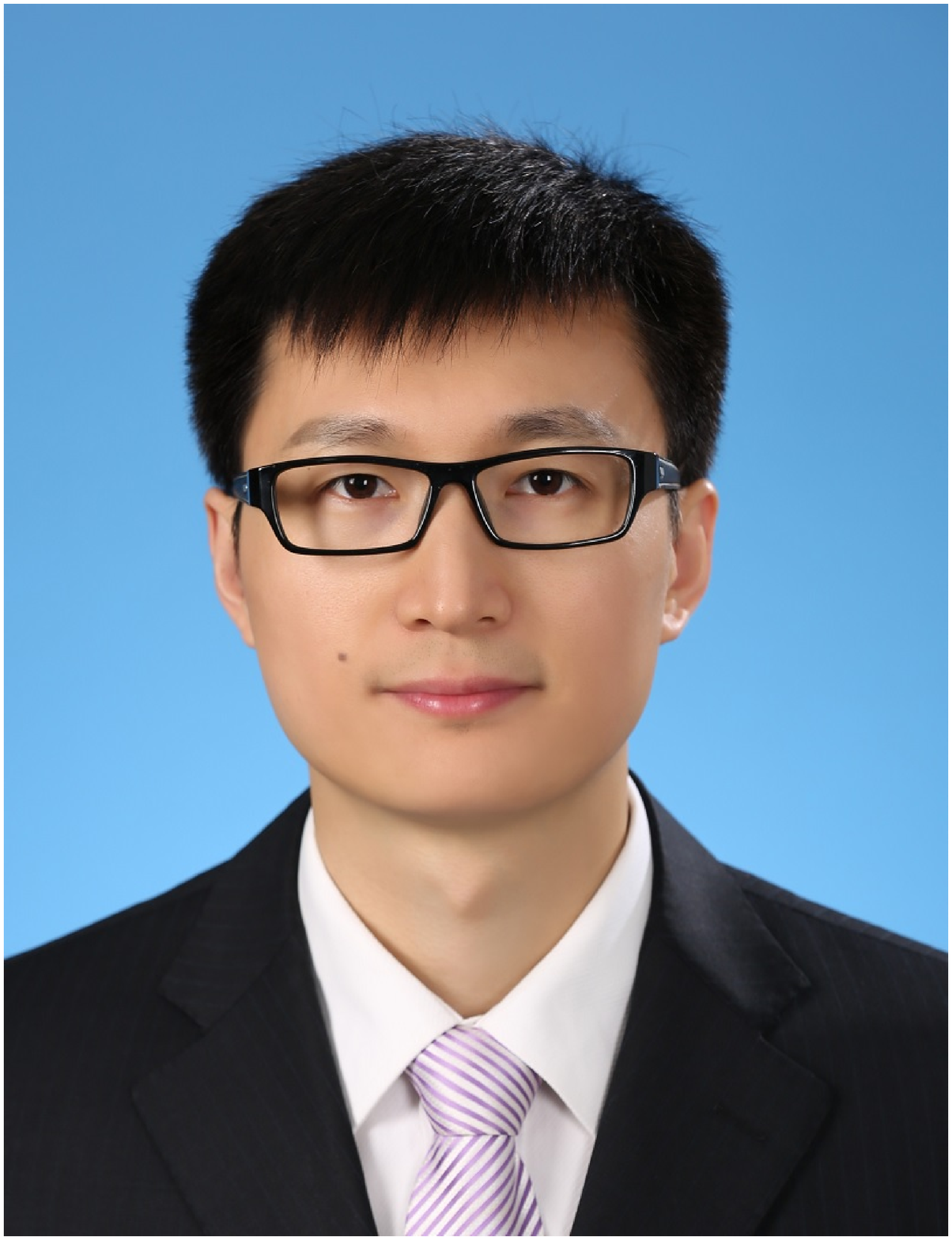}}]{Linglong Dai} (M'11-SM'14) received the B.S. degree from Zhejiang University, Hangzhou, China, in 2003, the M.S. degree (with the highest honor) from the China Academy of Telecommunications Technology, Beijing, China, in 2006, and the Ph.D. degree (with the highest honor) from Tsinghua University, Beijing, China, in 2011. From 2011 to 2013, he was a Postdoctoral Research Fellow with the Department of Electronic Engineering, Tsinghua University, where he was an Assistant Professor from 2013 to 2016 and has been an Associate Professor since 2016. His current research interests include reconfigurable intelligent surface (RIS), massive MIMO, millimeter-wave or Terahertz communications, and machine learning for wireless communications.

He has coauthored the book {\it MmWave Massive MIMO: A Paradigm for 5G} (Academic Press, 2016). He has authored or coauthored over 60 IEEE journal papers and over 40 IEEE conference papers. He also holds 19 granted patents. He has received five IEEE Best Paper Awards at the IEEE ICC 2013, the IEEE ICC 2014, the IEEE ICC 2017, the IEEE VTC 2017-Fall, and the IEEE ICC 2018. He has also received the Tsinghua University Outstanding Ph.D. Graduate Award in 2011, the Beijing Excellent Doctoral Dissertation Award in 2012, the China National Excellent Doctoral Dissertation Nomination Award in 2013, the URSI Young Scientist Award in 2014, the IEEE Transactions on Broadcasting Best Paper Award in 2015, the Electronics Letters Best Paper Award in 2016, the National Natural Science Foundation of China for Outstanding Young Scholars in 2017, the IEEE ComSoc Asia-Pacific Outstanding Young Researcher Award in 2017, the IEEE ComSoc Asia-Pacific Outstanding Paper Award in 2018, the China Communications Best Paper Award in 2019, the IEEE Access Best Multimedia Award in 2020, and the IEEE Communications Society Leonard G. Abraham Prize in 2020. He was listed as a Highly Cited Researcher by Clarivate Analytics in 2020.

He is an Area Editor of {\sc IEEE Communications Letters}, and an Editor of {\sc IEEE Transactions on Communications} and {\sc IEEE Transactions on Vehicular Technology}. Particularly, he is dedicated to reproducible research and has made a large amount of simulation codes publicly available.
\end{IEEEbiography}

\begin{IEEEbiography}[{\includegraphics[width=1in,height=1.25in,clip,keepaspectratio]{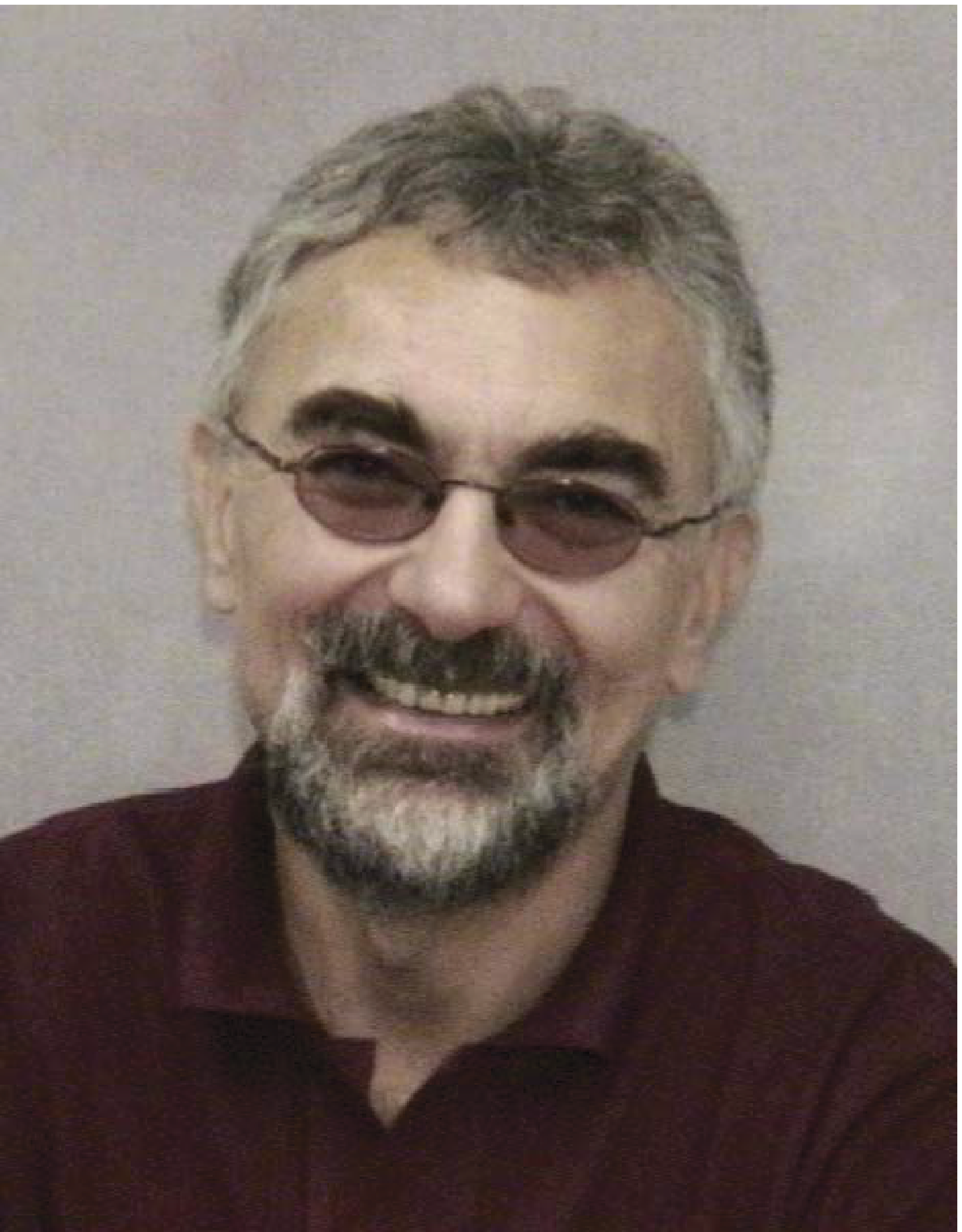}}]{Lajos Hanzo}
(\url{http://www-mobile.ecs.soton.ac.uk}, \url{https://en.wikipedia.org/wiki/Lajos_Hanzo}) (FIEEE'04) received his 
Master degree and Doctorate in 1976 and 1983, respectively from the 
Technical University (TU) of Budapest. He was also awarded the Doctor of 
Sciences (DSc) degree by the University of Southampton (2004) and 
Honorary Doctorates by the TU of Budapest (2009) and by the University 
of Edinburgh (2015).  He is a Foreign Member of the Hungarian Academy of 
Sciences and a former Editor-in-Chief of the IEEE Press.  He has served 
several terms as Governor of both IEEE ComSoc and of VTS.  He is also a 
Fellow of the Royal Academy of Engineering (FREng), of the IET and of 
EURASIP.
\end{IEEEbiography}

\end{document}